\documentclass[3p]{elsarticle}
\pdfoutput=1
\usepackage[latin1]{inputenc}
\usepackage{graphicx}
\usepackage{hyperref}
\usepackage{upgreek}
\usepackage{lineno}
\usepackage{amssymb}
\usepackage{array}

\journal{Journal of Instrumentation}

\biboptions{number}

\begin{document}

\title{Simulation study of the impact of AGIPD design choices on X-ray Photon Correlation Spectroscopy utilizing the intensity autocorrelation technique}

\author[hh]{Julian Becker\corref{cor1}}
\ead{Julian.Becker@desy.de}
\author[hh]{Christian Gutt}
\author[hh]{Heinz Graafsma}
\author[]{on behalf of the AGIPD consortium}
\cortext[cor1]{Corresponding author}
\address[hh]{Deutsches Elektronen-Synchrotron,\\ Notkestr. 85, 22607 Hamburg, Germany}

\begin{abstract}
The European XFEL, currently under construction, will produce a coherent X-ray pulse every 222~ns in pulse trains of up to 2700 pulses. In conjunction with the fast 2D area detectors currently under development, it will be possible to perform X-ray Photon Correlation Spectroscopy (XPCS) experiments on sub-microsecond timescales with non-ergodic systems.

A case study for the Adaptive Gain Integrating Pixel Detector (AGIPD) at the European XFEL employing the intensity autocorrelation technique was performed using the detector simulation tool HORUS. As optimum results from XPCS experiments are obtained when the pixel size approximates the (small) speckle size, the presented study compares the AGIPD (pixel size of (200~$\upmu$m)$^2$) to a possible apertured version of the detector and to a hypothetical system with (100~$\upmu$m)$^2$ pixel size and investigates the influence of intensity fluctuations and incoherent noise on the quality of the acquired data.

The intuitive conclusion that aperturing is not beneficial as data is 'thrown away' was proven to be correct for low intensities. For intensities larger than approximately 1 photon per (100~$\upmu$m)$^2$ aperturing was found to be beneficial, as charge sharing effects were excluded by it.

It was shown that for the investigated case (100~$\upmu$m)$^2$ pixels produced significantly better results than (200~$\upmu$m)$^2$ pixels when the average intensity exceeded approximately 0.05 photons per (100~$\upmu$m)$^2$.

Although the systems were quite different in design they varied in the signal to noise ratio only by a factor of 2-3, and even less in the relative error of the extracted correlation constants. However the dependence on intensity showed distinctively different features for the different systems.

\end{abstract}

\begin{keyword}
AGIPD \sep simulation \sep XFEL \sep X-ray Photon Correlation Spectroscopy \sep aperturing \sep intensity fluctuations
\end{keyword}

\maketitle

\linenumbers

\section{Introduction}

The European X-Ray Free Electron Laser (XFEL) \cite{XFEL} will provide ultra short, highly coherent X-ray pulses which will revolutionize scientific experiments in a variety of disciplines spanning physics, chemistry, materials science and biology. The European XFEL will provide pulse trains of up to 2700 pulses every 222~ns (600~$\upmu$s in total) followed by an idle time of 99.4~ms, resulting in a supercycle of 10~Hz and 27000 pulses per second.

One of the differences between the European XFEL and other free electron laser sources is the fast pulse repetition frequency of 4.5~MHz. Dedicated fast 2d detectors are being developed, one of which is the Adaptive Gain Integrating Pixel Detector (AGIPD) \cite{AGIPD1, AGIPD2, AGIPD3}. The development is a collaboration between DESY, the University of Hamburg, the University of Bonn (all in Germany) and the Paul Scherrer Institute (PSI) in Switzerland.

The technique of X-ray Photon Correlation Spectroscopy (XPCS) provides an experimental method to probe dynamic properties of condensed matter \cite{grubel} on nanometer length-scales. In XPCS experiments samples are illuminated by coherent X-ray light, which results in a grainy diffraction pattern (speckle pattern). As the speckle pattern is determined by the exact spatial arrangement of all the particles in the beam, any motion of the particles results in a fluctuation of the speckle pattern. The basic quantity determined in XPCS experiments is the intensity autocorrelation function of the speckles which measures tiny density fluctuations in the sample as a function of time and scattering vector $Q$. The high pulse repetition rate of the European XFEL, combined with the fast frame rate of the AGIPD, will allow investigations using the XPCS technique on the sub-microsecond timescale.

Depending on the dynamic phenomena of interest and the associated time scales different technical realizations of photon correlation spectroscopy exist. For the very fast time scales in the regime of $10^{-12}$ to $10^{-9}$ seconds, the split and delay technique can be applied \cite{split_delay1, split_delay2, gutt}. To study effects on the time scale of $10^{-6}$ to $10^{0}$ seconds, which are relevant in soft matter or biological systems, sequential XPCS is performed \cite{grubel}.

Photon correlation experiments are limited by their signal to noise ratio (S/N or SNR) \cite{jakeman}. An important aspect of the SNR in XPCS experiments is the ability of the X-ray detector to resolve the speckle pattern. It can be shown that an optimal signal to noise ratio is achieved in situations where the speckle size equals the pixel size \cite{falus}. As XPCS seeks to measure dynamic sample properties, the X-ray pulses must illuminate the same sample spot several times. This implies that sample distortions (especially sample heating) due to interactions with the illuminating beam should be reduced as much as possible by using a large beam size. The speckle size is inversely proportional to the beam size and therefore speckle sizes will be small for XPCS experiments at XFEL sources. The pixel sizes of the fast 2D-detectors under development will be larger than the expected speckle sizes. 
This study investigates to what extent this mismatch will affect the SNR and the quality of the intensity correlation functions. Simulations of a prototypical XPCS experiment performed for different detector systems in otherwise identical situations are presented. The study is limited to small angle scattering experiments and shows the influence of the detector system on the SNR, which is commonly neglected.

Furthermore, this study investigates a method of reducing the effective pixel size by aperturing the sensitive area of each pixel. Aperturing pixels produces a detector consisting of smaller pixels, that are isolated and separated from each other. Compared to adjacent small pixels, this eliminates charge sharing and makes it possible to sample a larger area with the same number of pixels. On the other hand the fraction of the detector area sensitive to photons is greatly reduced.

This study is the first to investigate the performance of the AGIPD and the influence of certain design choices on the data quality of the experiment. A system level simulation study using an XPCS case for the DSSC\footnote{The DEPFET Sensor with Signal Compression (DSSC) project and Large Pixel Detector (LPD) project are the other two projects developing a dedicated detector for experiments at the European XFEL.} project can be found in literature \cite{hansen}. 

\section{Simulation of a prototypical XPCS experiment at the European XFEL}

\begin{figure}[tb]
	\centering
		\includegraphics[width=0.8\textwidth]{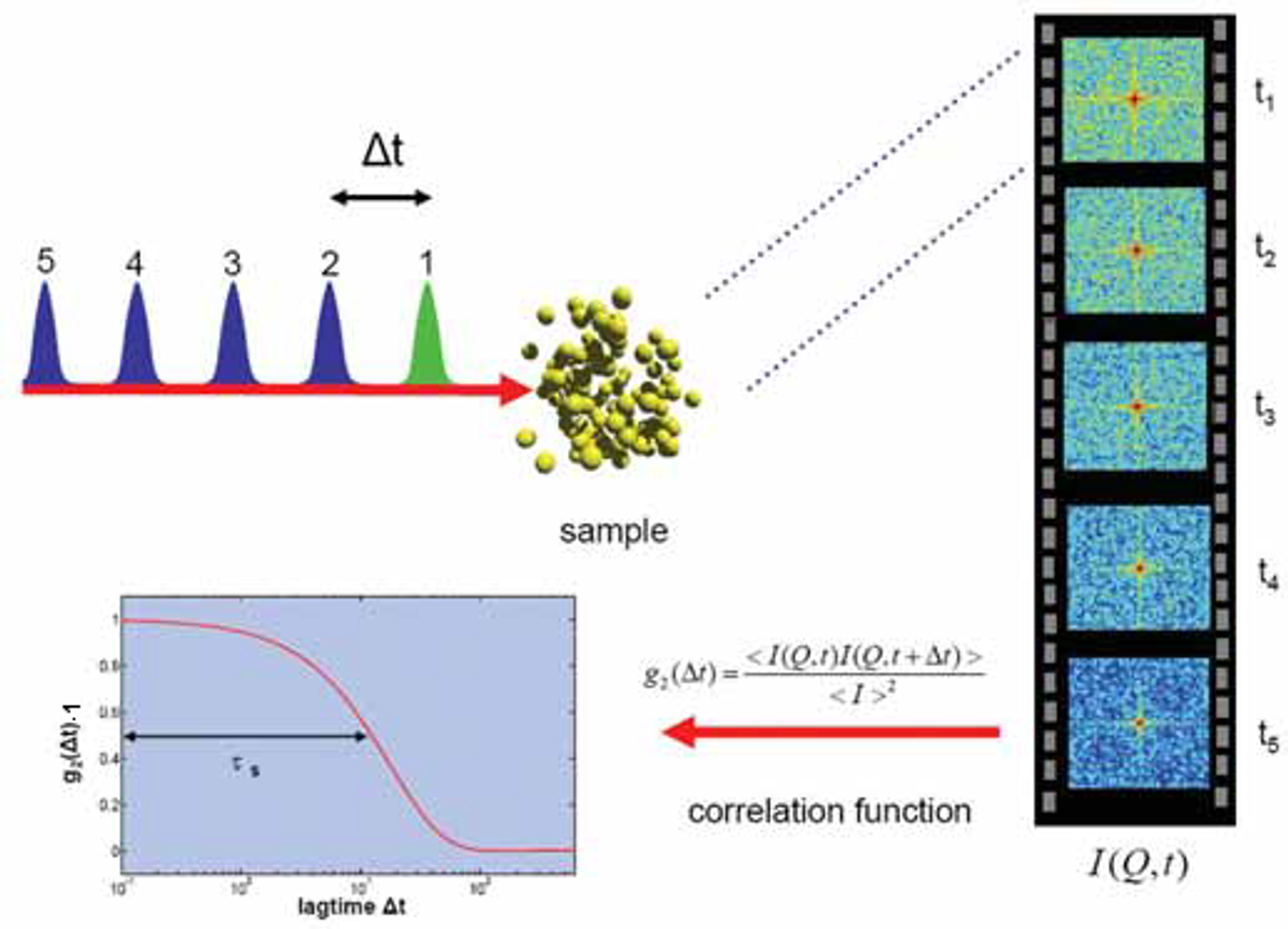}
	\caption{Illustration of an exemplary XPCS experiment. Typical distances between sample and detector for experiments at the European XFEL are about 10~m. The grains on the diffraction patterns are speckles. The images used in this study differ from the shown image, as explained later in this manuscript. Image reproduced from the European XFEL TDR \cite{XFEL}.}
	\label{setup}
\end{figure}

All the simulations have been performed within the IDL framework\footnote{IDL stands for Interactive Data Language and is distributed by ITT Visual Information Solutions. For more information visit \url{http://www.ittvis.com/}} version 7.1. The simulations consist of three independent stages modeling the real space system, the diffraction image generation and the detector response. The stages are explained in detail below. The simulations are followed by a data evaluation step, which is explained in section \ref{data_eval}. 

The simulations use a photon energy of 12.4~keV ($\lambda=1$~\AA), which is the design energy of the European XFEL. Most XPCS experiments will be conducted using the Materials Imaging and Dynamics (MID) experimental station, where the distance between sample and detector will be around 10~m. An illustration is shown in Figure \ref{setup}.

\subsection{Real space system}

\begin{figure}[tb]
	\centering
		\includegraphics[width=0.35\textwidth]{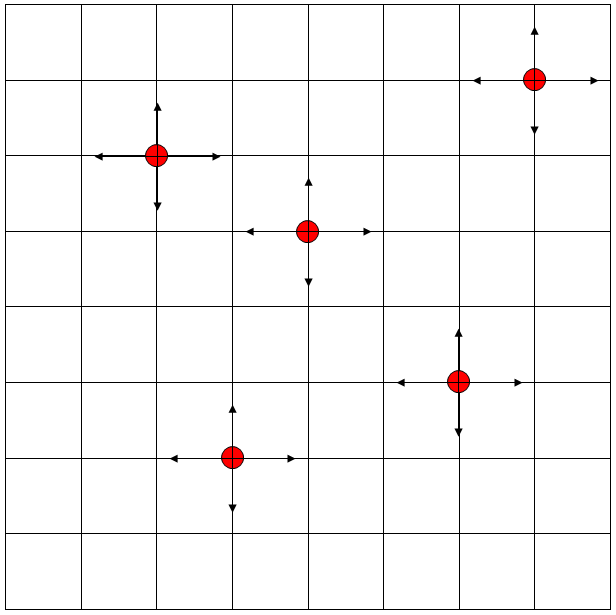}
	\caption{Illustration of the real space system. The particles did not interact with each other and could occupy the same grid position. Per simulation step each particle could move by $\pm1$ grid position in each of the dimensions independently. Particles leaving the grid on one side were replaced by particles entering on the opposite side. In total 5000 particles were simulated on a square grid of 2000$\times$2000 points. The initial distribution of particles was random.}
	\label{2d_grid}
\end{figure}

The simulated diffraction patterns were based on the evolution of a real space system. For ease of simulation the simulated real space system consisted of particles hopping on a two dimensional grid as illustrated in Figure \ref{2d_grid}. 
The particles did not interact with each other and could occupy the same grid position. Per simulation step ($\Delta t$) each particle could move by $\pm1$ grid position with a probability of 1~\% in each of the dimensions independently. Particles leaving the grid on one side were replaced by particles entering on the opposite side.
In total 5000 particles were simulated on a square grid of 2000$\times$2000 points. The initial distribution of particles was random.

\subsection{Generation of diffraction images}

For each simulation step the real space particle distribution was converted to an image. This real space image had the same size as the simulation grid and each image pixel was assigned the number of particles at the corresponding grid point. From this point onwards, all the calculations were done image-based.

\subsubsection{Speckle size, illumination function and Fourier transform}

\begin{figure}[tb]
	\centering
		\includegraphics[width=0.5\textwidth]{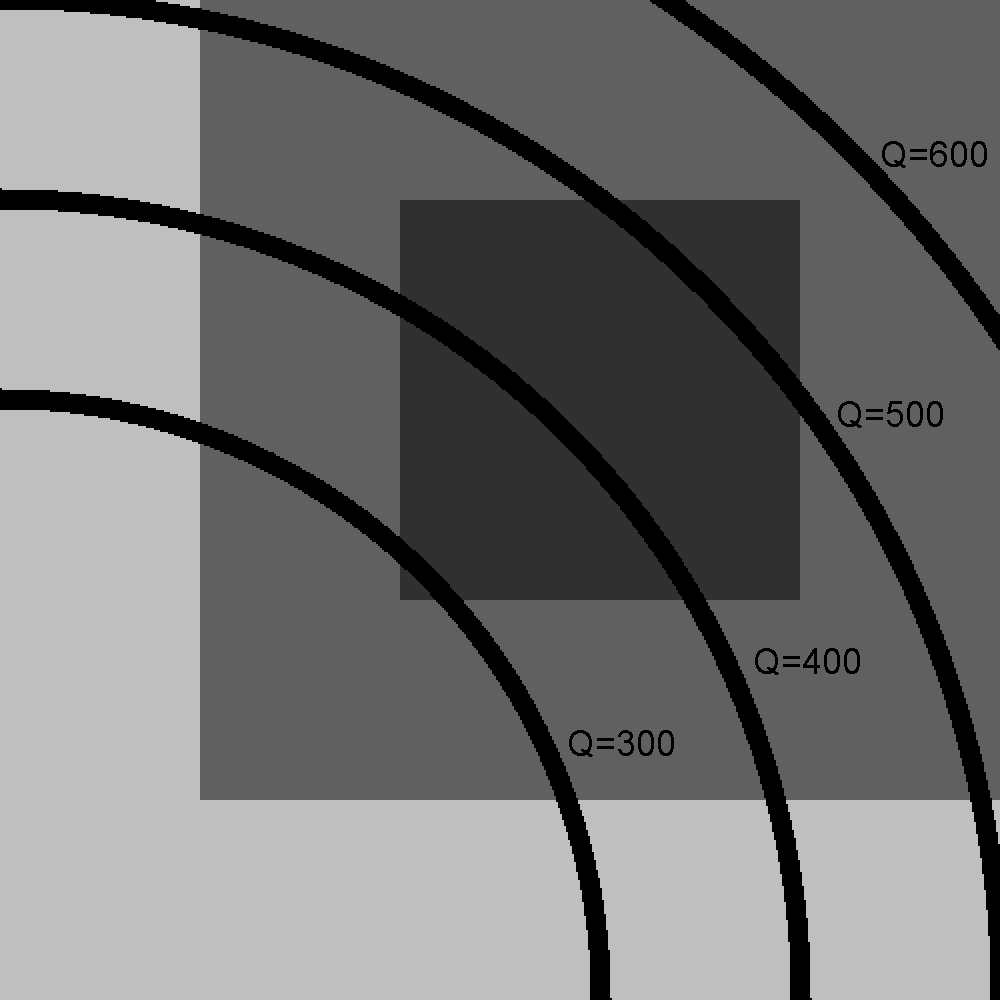}
	\caption{Regions of the diffraction pattern and their assignment in this study. Light gray: discarded pixels due to possible structure and form factors. Medium and dark gray: pixels used for all studies of (200~$\upmu$m)$^2$ detectors and small ROI investigations of (100~$\upmu$m)$^2$ detectors. Dark gray: pixels used for studies of (100~$\upmu$m)$^2$ detectors and large ROI investigations. Pixels assigned to the exemplary $Q$ bins of 300, 400, 500 and 600 units have been marked black.}
	\label{areas}
\end{figure}

\begin{figure}[tb!]
	\centering
	\begin{tabular}[t]{cc} 
		\includegraphics[width=0.56\textwidth]{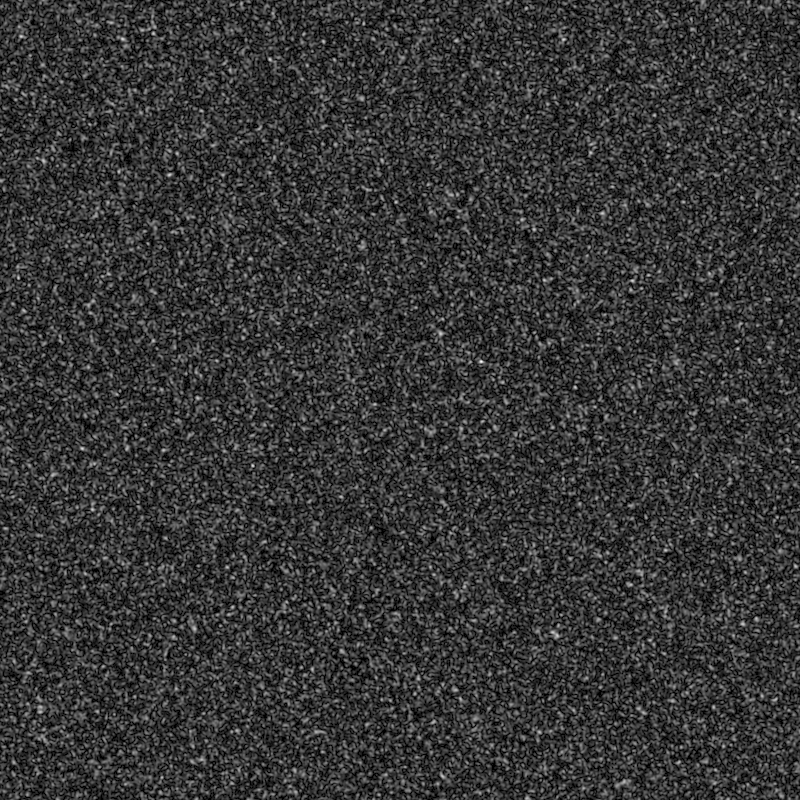} & \begin{tabular}[b]{c} 
		 \includegraphics[width=0.4\textwidth]{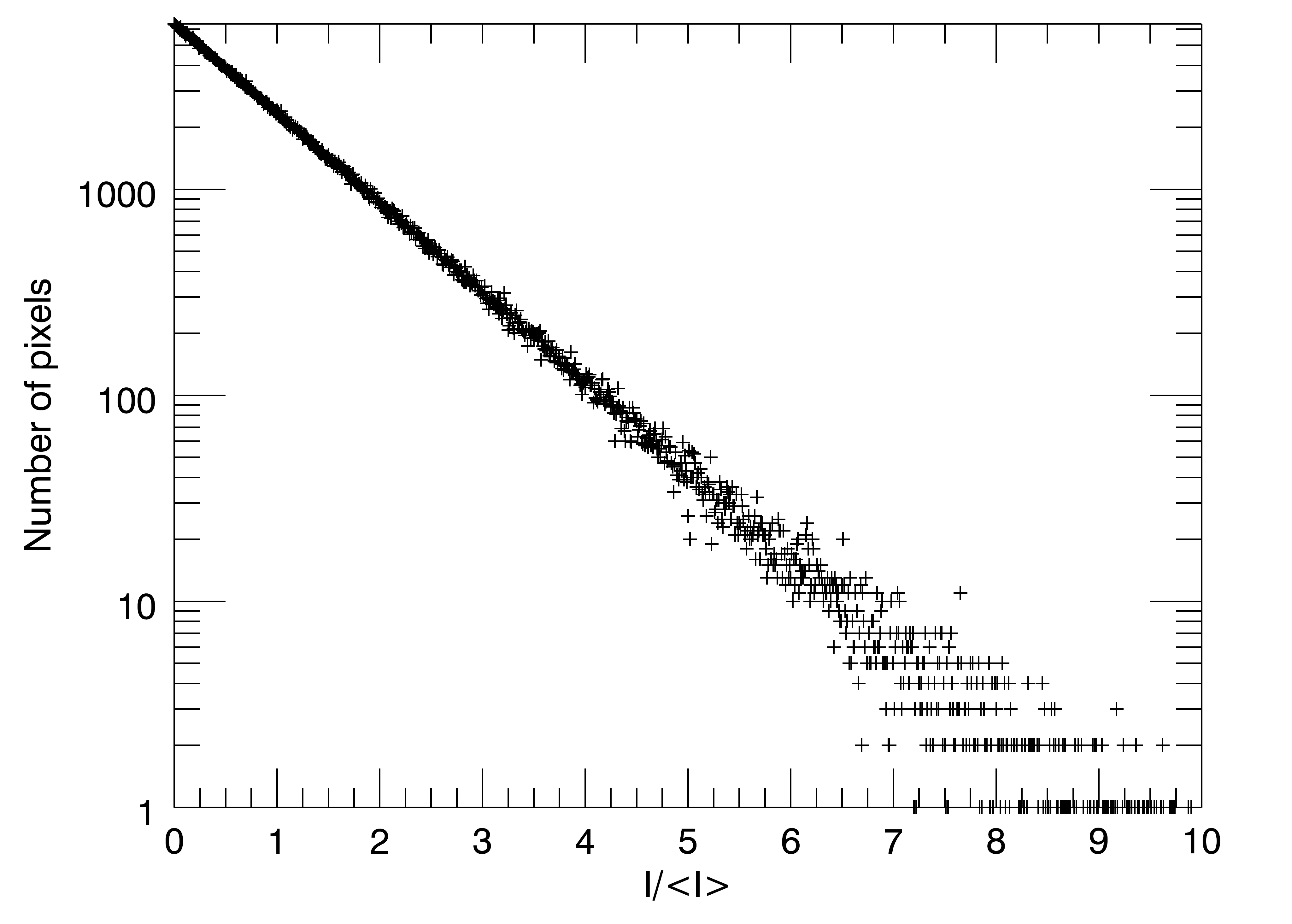} \\
		 \includegraphics[width=0.4\textwidth]{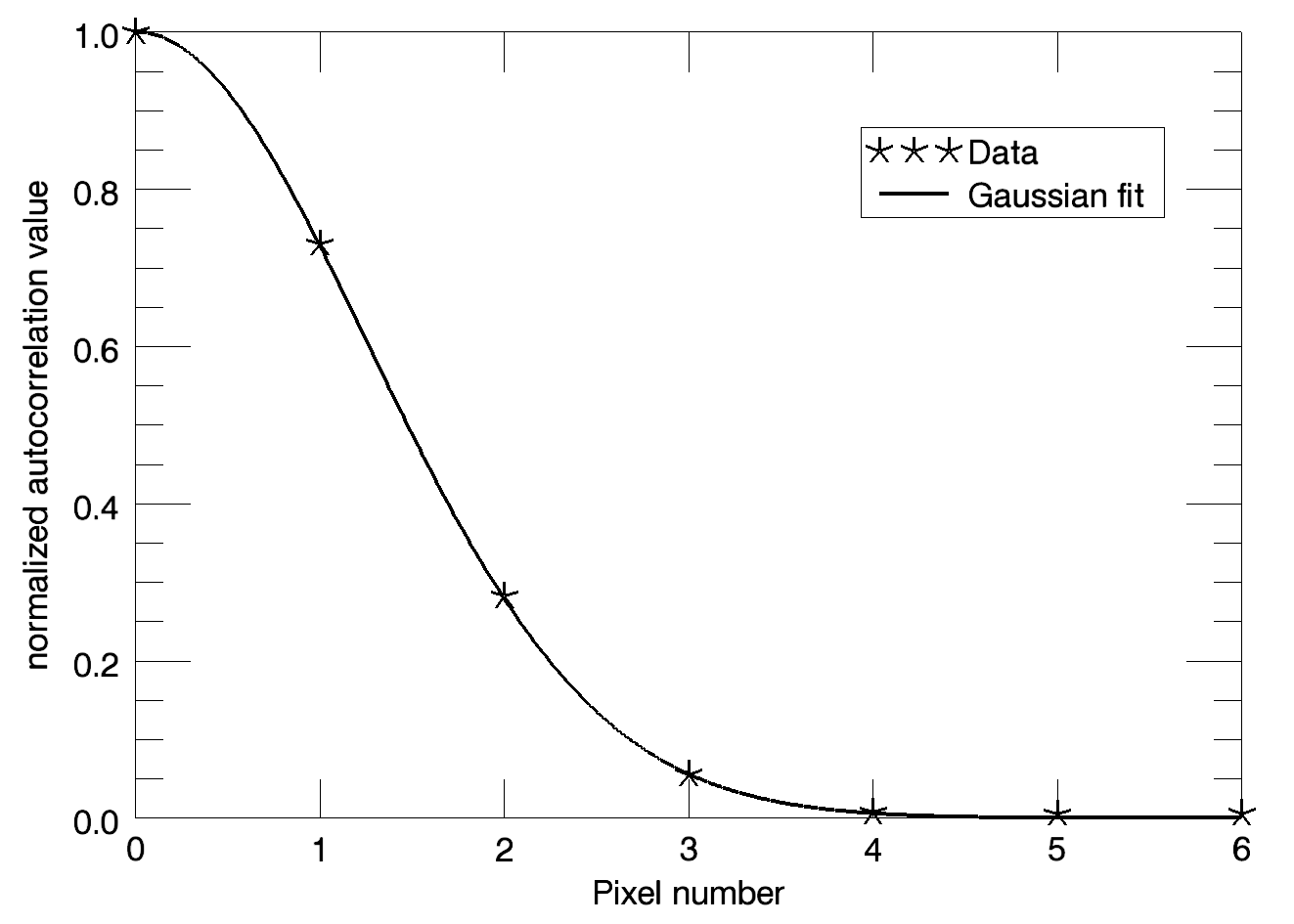} \\
		\end{tabular} \\
	\end{tabular}
	\caption{The left image shows the intensity distribution of one exemplary Fourier transform. No structure and form factor can be observed. The upper right image shows the histogram of the intensity distribution, binned with a bin size of $0.01 \langle  I\rangle  $. The lower right image shows the autocorrelation function along the x axis averaged over the y axis, thus revealing the average speckle shape. The speckle shape is excellently matched by a Gaussian fit with a sigma of 1.25 pixels.}
	\label{speckles}
\end{figure}

The speckle size S is usually given by the following expression:

\begin{equation}
	S \approx \frac{\lambda L}{D} \label{speckle_size}
\end{equation}
with $\lambda$ being the wavelength of the illuminating light, $L$ the sample to detector distance and $D$ the size of the illuminating light beam.
Therefore an illumination function, which corresponds to the spatial distribution of the primary beam, is needed  to define the speckle size in the detector plane. 

The complex wave field in the detector plane $E_{diff}$ was defined as:

\begin{equation}
	E_{diff} = FT\left\{ F_{rs} \ast F_{ill}\right\}
\end{equation}
where $F_{rs}$ is the real space image and $F_{ill}$ the illumination function with $FT$ denoting the discrete Fourier transform (FFT algorithm) and $\ast$ indicating pixel by pixel multiplication.
The illumination function was assumed to be a Gaussian function centered on the image center with a sigma corresponding to one eighth of the image size.

To remove the structural artifacts of the square image the first 200 pixels in x and y direction were discarded. The removed area is shown in light gray in Figure \ref{areas}.

A Fourier transform of real values is symmetric around the Nyquist frequency, as indicated by Friedel's Law. As $E_{diff}(Q)=E_{diff}^*(-Q)$, areas symmetric to Figure \ref{areas} were removed from the complex wave field, leaving data from a region of 800$\times$800 pixels for further processing, shown in medium and dark gray in Figure \ref{areas}. The intensities from an exemplary wave field are shown in the left image of Figure \ref{speckles} and exhibit no structure and form factor effects.

The upper right image of Figure \ref{speckles} shows a histogram of the intensities of the complex wave field with a bin size of one hundredth of the average intensity {$\langle  I\rangle  $. The expected negative exponential distribution \cite{xpcs_scm_book} for fully coherent illumination is clearly reproduced and the most probable intensity value is zero. The large variation in the histogram bins at large values of \mbox{$I/\langle  I\rangle  $} is due to the finite number of image pixels.

The lower right image of Figure \ref{speckles} shows the average normalized spatial autocorrelation function $ACF$ of the intensities along the x direction averaged over the y direction defined by:

\begin{equation}
	ACF(\Delta R) = \frac{1}{N_y}\sum_{l=0}^{N_y-1}{\frac{\sum_{k=0}^{N_x-\Delta R-1} {(I_{k,l}-\langle  I\rangle  _l)(I_{k+\Delta R,l}-\langle  I\rangle  _l)}} {\sum_{k=0}^{N_x-1}{(I_{k,l}-\langle  I\rangle  _l)^2}}}
\end{equation}
where $\Delta R$ is the inter pixel distance, $N_x$ and $N_y$ are the number of pixels along the corresponding direction, $I_{k,l}$ the intensity in the indicated pixel and $\langle  I\rangle  _l$ the average intensity in the corresponding line. 

By calculating the average normalized spatial autocorrelation function the speckle shape and size was determined independently from the knowledge of the illumination function. The speckle shape corresponds excellently to the Gaussian fit with a sigma of 1.25 pixels\footnote{As the Fourier transform is oversampled and one pixel length corresponds to 50~$\upmu$m, this results in a sigma of 62.5~$\upmu$m and a full width at half maximum of 147.2~$\upmu$m for the speckles in this simulation study. Assuming 10~m distance between sample and detector, this yields a FWHM of 6.8~$\upmu$m of the illuminating beam.}. Exchanging x and y direction in the evaluation yielded identical results, indicating symmetric speckles.

Speckle positions can be misaligned with the detector pixel grid. In order to simulate this, 2 by 2 pixels of the Fourier transform were assigned to each (100~$\upmu$m)$^2$ pixel area. Correspondingly (200~$\upmu$m)$^2$ area were assigned 4 by 4 pixels of the complex wave field. The sub-pixel position information gained by this procedure was used in the detector simulation step.

\subsubsection{Intensity normalization and fluctuation, image quantization and incoherent noise}
\label{image}

The diffraction pattern impinging on the detector consists of discrete photons. This discrete photon distribution $F_\gamma$ was generated in the following way:

\begin{equation}
	F_\gamma = S\left\{ \langle  I\rangle  _{d} R_{fluct} \frac{I(E_{diff})}{\langle  I(E_{diff})\rangle  } \right\}+ F_{noise} \label{quant}
\end{equation}
where $S(X)$ is the image quantization procedure explained below, $\langle  I\rangle  _{d}$ the desired average intensity, $R_{fluct}$ the intensity fluctuation factor explained below, $I(E_{diff})=\left|E_{diff}\right|^2$ the intensity distribution in the detector plane, $\langle  I(E_{diff})\rangle  $ the average intensity in the detector plane and $F_{noise}$ the additional incoherent noise explained below.

The fully coherent wave field is ergodic\footnote{In very simplified terms: An ergodic system shows the same average behavior when probed an infinite number of times sequentially as an infinite number of identical systems would show when probed once.}. In order to quantize the photon field (denoted by $S(X)$), the photons were sampled according to the negative binomial distribution expected for its spatial distribution \cite{xpcs_scm_book}, using the floating point pixel intensity as the expected value $\langle  I\rangle  $:

\begin{equation}
	P(I)=\frac{\Gamma(I+M)}{\Gamma(M)\Gamma(I+1)} \left(1+\frac{M}{\langle  I\rangle  }\right)^{-I} \left(1+\frac{\langle  I\rangle  }{M}\right)^{-M} \label{p_dist_part}
\end{equation}
where $I$ is a discrete number of photons, $P(I)$ the probability of realizing $I$ photons and $M$ the number of modes in the image. $\Gamma(x)=(x-1)!$ denotes the gamma function. Thus, for the fully coherent case ($M=1$) investigated in this study, equation \ref{p_dist_part} simplifies to:

\begin{equation}
	P(I)=\frac{1}{1+\langle  I\rangle  } \left(\frac{\langle  I\rangle  }{1+\langle  I\rangle  }\right)^{I} \label{p_dist}
\end{equation}

The intensity fluctuation factor $R_{fluct}$ was either set to unity (no fluctuations) or sampled randomly from a Gaussian distribution with a mean of 1 and an rms of 0.1. The factor does not simulate intensity fluctuations directly, but rather remaining uncertainties after accounting for intensity fluctuations. Intensity fluctuations can be accounted for in the determination of the autocorrelation function by weighting images, however the most probable value of zero photons cannot be scaled, hence creating uncertainties. Accounting for intensity fluctuations is particularly important at FEL sources, as the pulse to pulse variations can be quite large \cite{fel_fluct}.

The incoherent noise $F_{noise}$ added after the quantization process was either zero or considered Poissonian with a probability of $1/400$ per pixel of the Fourier transform. This translates to a noise photon probability of $1/100$ per (100~$\upmu$m)$^2$ detector area. Thus for an average intensity of 0.01 the SNR from the experiment itself is 1. As the noise of the detection process was simulated separately (see below) this completely random noise can be considered a high estimate of fluctuations of an otherwise constant background distribution.

\subsection{Detector response}

In order to evaluate the effects of the investigated detector systems (detailed below), the same photon distributions $F_\gamma$ were taken as input for the different detector systems.

\subsubsection{Simulated detector systems}

To evaluate the influence of different detector choices the following detector systems were simulated:

\begin{itemize}
	\item Ideal detector systems with (100~$\upmu$m)$^2$ and (200~$\upmu$m)$^2$ pixel size. 
	\item The Adaptive Gain Integrating Pixel Detector (AGIPD) featuring (200~$\upmu$m)$^2$ pixel size.
	\item A Modified AGIPD using Aperturing Techniques (MAAT), which was identical to the AGIPD, but completely insensitive to photons which are closer than 50~$\upmu$m to a pixel boundary\footnote{For real applications this could e.g. be realized by post-processing a tungsten grid of 10~$\upmu$m thickness on the entry window of the detector.}, which resulted in an effective pixel size of (100~$\upmu$m)$^2$.
	\item A Reduced Amplitude SEnsing System (RAMSES), which was a hypothetical detector which sacrificed most of the AGIPD functionality (e.g. the gain switching) but featured (100~$\upmu$m)$^2$ pixel size.
\end{itemize}

\subsubsection{The Adaptive Gain Integrating Pixel Detector (AGIPD)}
AGIPD is based on the hybrid pixel technology. A newly developed Application Specific Integrated Circuit (ASIC) will feature in each pixel a dynamic gain switching amplifier (to cope with the high dynamic range) and an analogue memory capable of storing approximately 300 images at the desired 4.5~MHz speed. 

The AGIPD will feature a pixel size of (200~$\upmu$m)$^2$ and a silicon sensor with a thickness of 500~$\upmu$m. The area of (200~$\upmu$m)$^2$ is needed to realize in-pixel electronics which fulfill the requirements for detectors at the European XFEL \cite{AGIPD1, AGIPD2}. The image data is read out and digitized in the gap between two pulse trains.

As all the photons arrive at the detector in a period of time that is very short compared to the time needed for charge transport in the sensor material, the AGIPD finds the number of photons absorbed in each pixel by integrating the total signal. 

The AGIPD detector will be available for the first experiments at the European XFEL. The other systems studied here have been suggested as possible modifications for XPCS experiments. Preliminary studies showed that a system with reduced pixel size (RAMSES) would not be able to store as many frames as the AGIPD; maybe fewer than 100.

\subsubsection{Physical constraints precluding the realization of an ideal detector}

Due to the special pulse structure of the European XFEL, it is necessary to store the acquired images inside the pixel logic during the pulse train. A compromise has to be found between a large pixel area, so that many images can be stored, and a small pixel area for high spatial resolution.

Additionally, effects of the limited stopping power of silicon and the charge transport in the sensitive volume come into play. The chosen sensor thickness of 500~$\upmu$m provides a quantum efficiency exceeding 85\% for 12~keV photons. After a photon is absorbed, a cloud of charge carriers is created, which expands before reaching the readout electrode and may cause charge sharing between pixels. The number of electron hole pairs created per absorption is normally distributed with a mean of $\langle  N_{e,h}\rangle  =E_\gamma/3.6~eV$ and a variance of $\sigma_{e,h}^2=F\langle  N_{e,h}\rangle  $, with $F\approx 0.1$ being the Fano factor of silicon.

The actual shape and size of this cloud is influenced by many parameters\footnote{A non-exhaustive list includes: bias voltage, sensor doping, number of charge carriers, electrode geometry and exact initial position.}. 
In this study, the collected charge is approximated as following a Gaussian distribution with a width depending of the square root of the distance to the electrode \cite{fowler}.
To minimize charge sharing between the pixels, the charge cloud size should be small compared to the pixel size\footnote{For XPCS applications charge sharing should be avoided if possible, as the signal of adjacent pixels would be correlated, increasing the effective pixel size. It should be noted that for very low intensities (low pixel occupancies) charge sharing is beneficial, as event-by-event correlations can be performed \cite{droplet}.}. Given the material properties of a typical silicon detector of 500~$\upmu$m thickness and using recent determinations of the charge carrier mobilities \cite{sse}, typical cloud sizes (full width at half maximum) are in the range of 20~$\upmu$m\footnote{200~V sensor bias results in drift times $t_d$ around 30~ns, assuming the whole thickness as drift distance. The lateral cloud size can then be approximated as $FWHM_{cloud}\approx 2.35\sqrt{2Dt_d}$ and $D=\mu kT/e$ with $\mu , k, T, e$ being hole mobility, Boltzmann constant, temperature and electron charge, respectively. The FWHM value of 20~$\upmu$m is the result of assuming $\langle 100 \rangle$ crystal orientation and a temperature of 300~K.}. 

\subsubsection{HORUS}

The detector response was calculated using the HORUS software described in \cite{AGIPD3, horus1}. HORUS has already been successfully used to simulate the performance of the Medipix3 chip \cite{david1, david2} and recently for AGIPD \cite{horus3}. The entry window of the sensor was included as 3~$\upmu$m of insensitive silicon equivalent material for all systems, which is about twice the thickness foreseen for AGIPD at the moment. For all systems, the total noise was assumed to have an equivalent noise charge of approximately 300 electrons, which is about double the value determined from ASIC simulations \cite{AGIPD2} and roughly the value measured on a test assembly using a microstrip sensor. 

To increase the simulation speed several approximations were used:

\begin{itemize}
	\item Any parallax effects were ignored. This is a reasonable approximation, as the sample to detector distance for XPCS experiments at the European XFEL will probably be 10~m or more.
	\item Any effects originating from sensor edges and module gaps were ignored. Although in the final detector system these effects cannot be avoided, for XPCS data evaluation it is possible to mask areas exhibiting these effects and exclude them from the data analysis.
	\item Effects produced by high instantaneous charge densities (so called plasma effects) are not simulated by HORUS, and thus were not included in this study. Neglecting these effects is a reasonable approximation, as the intensities in XPCS experiments are usually low and plasma effects are expected to show only for intensities larger than 250 photons per pulse per (100~$\upmu$m)$^2$ area \cite{plasma, thesis, plasma2}.
\end{itemize}

The ideal detectors were realized by taking the total number of photons impinging on each pixel. Thus there is no detector noise and the quantum efficiency of the ideal systems is 1.0 (in contrast to $\approx 0.89$ for the non-ideal systems).

The response of the MAAT was calculated in an identical way as for the AGIPD, but photons hitting the insensitive regions were discarded in $F_\gamma$.

The RAMSES was modeled using the AGIPD parameters, except that the pixel size in the simulations was (100~$\upmu$m)$^2$, quadrupling the number of pixels in the simulation (as the same $Q$ space is covered).

\subsubsection{Region of interest}

\begin{figure}[tb]
	\centering
		\includegraphics[width=0.8\textwidth]{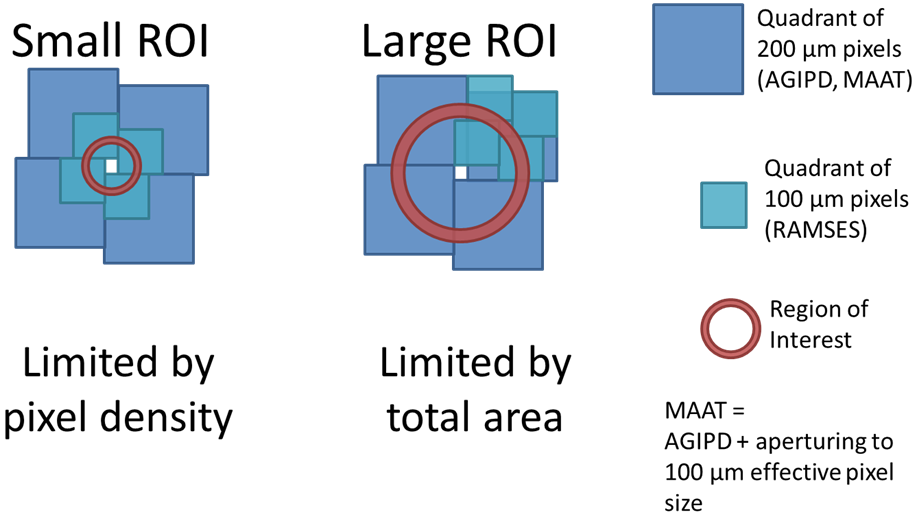}
	\caption{Illustration emphasizing the difference between small and large areas of interest. The actual pixels simulated and their assignment to the regions of interest is explained in detail in Figure \ref{areas}.}
	\label{roi}
\end{figure}

There will be experimental cases where the entire region of interest can be sampled with the detector area. For these cases the detector performance is limited by the pixel density, and similar total areas covered by (100~$\upmu$m)$^2$ and (200~$\upmu$m)$^2$ pixels had to be compared. These cases are labeled as cases with small Region Of Interest (sROI) in the rest of this study. An illustration of this is shown in Figure \ref{roi}.

For cases where the potential region of interest exceeds the detector area, (100~$\upmu$m)$^2$ and (200~$\upmu$m)$^2$ systems with an identical number of pixels had to be compared.  The $Q$ region for the simulated small pixel detectors was limited to the same number of pixels that was used for the large pixel detectors, shown as the dark gray area in Figure \ref{areas}. These cases are labeled as cases with large Region Of Interest (lROI).

It should be noted that, translated to real physical systems, the $Q$ vector units in the small and large ROI cases would therefore have different meanings.

\section{Data evaluation}
\label{data_eval}

The simulation was performed multiple times for each intensity, simulating the data acquisition of five independent\footnote{In the simulations each train used a new random number sequence.} pulse trains ($N_{tr}=5$), thus the overall data set consisted of 5~$\times$~300 frames ($N_f=300$) per detector type per intensity point.

\subsection{The intensity autocorrelation function}

The intensity autocorrelation function  ($g_2$ function) is both a measure of the correlation time and the contrast. For ideal circumstances, which are not realized here, the $g_2$ function would be a function monotonically decreasing from a value of 2 for $k\Delta t \rightarrow 0 $ to a value of 1 for $k\Delta t \rightarrow \infty $. The functional form is determined by the sample dynamics and the correlation time constant can be extracted from this. The scaling is a measure of the optical contrast.

For the data evaluation the $g_2$ function was calculated individually for each pixel p according to the following formula\footnote{The shown equations use the commonly employed normalization scheme. The data was also analyzed using the symmetric normalization scheme discussed in \cite{schaetzel}, however the results were worse then for the commonly employed normalization scheme and are therefore not shown here.}:

\begin{equation}
	g_2\left(p,k\Delta t\right) = \frac{1}{N_{tr}}\sum_{tr=0}^{N_{tr}-1}{\frac{1}{N_f-k}\sum^{N_f-k-1}_{i=0}{\frac{I_{p,tr}\left(i\Delta t\right) I_{p,tr}\left((i+k)\Delta t\right)}{\langle  I_p\rangle  ^2}} } \label{IAK}
\end{equation}
where the average intensity of pixel p ($\langle  I_p\rangle  $) is calculated as: 
\begin{equation}
	\langle  I_p\rangle   = \frac{1}{N_{tr}}\sum_{tr=0}^{N_{tr}-1}{\frac{1}{N_f}\sum^{N_f-1}_{i=0}{I_{p,tr}\left(i\Delta t\right)}}
\end{equation}
and $\Delta t$ is the time difference between frames and $I_{p,tr}\left(i\Delta t\right)$ denotes the intensity in pixel $p$ in frame $i$ of pulse train $tr$. This procedure yielded a $g_2$ function for each pixel of each investigated average intensity and detector type. 

Especially at very low intensities, there were pixels which did not detect any photon in the whole series. To avoid division by zero (as $\langle  I_p\rangle  =0$) the $g_2$ function for these pixels was set to zero.

\subsection{$Q$ space binning and azimuthal average}

\begin{figure}[tb]
	\centering
		\includegraphics[width=0.6\textwidth]{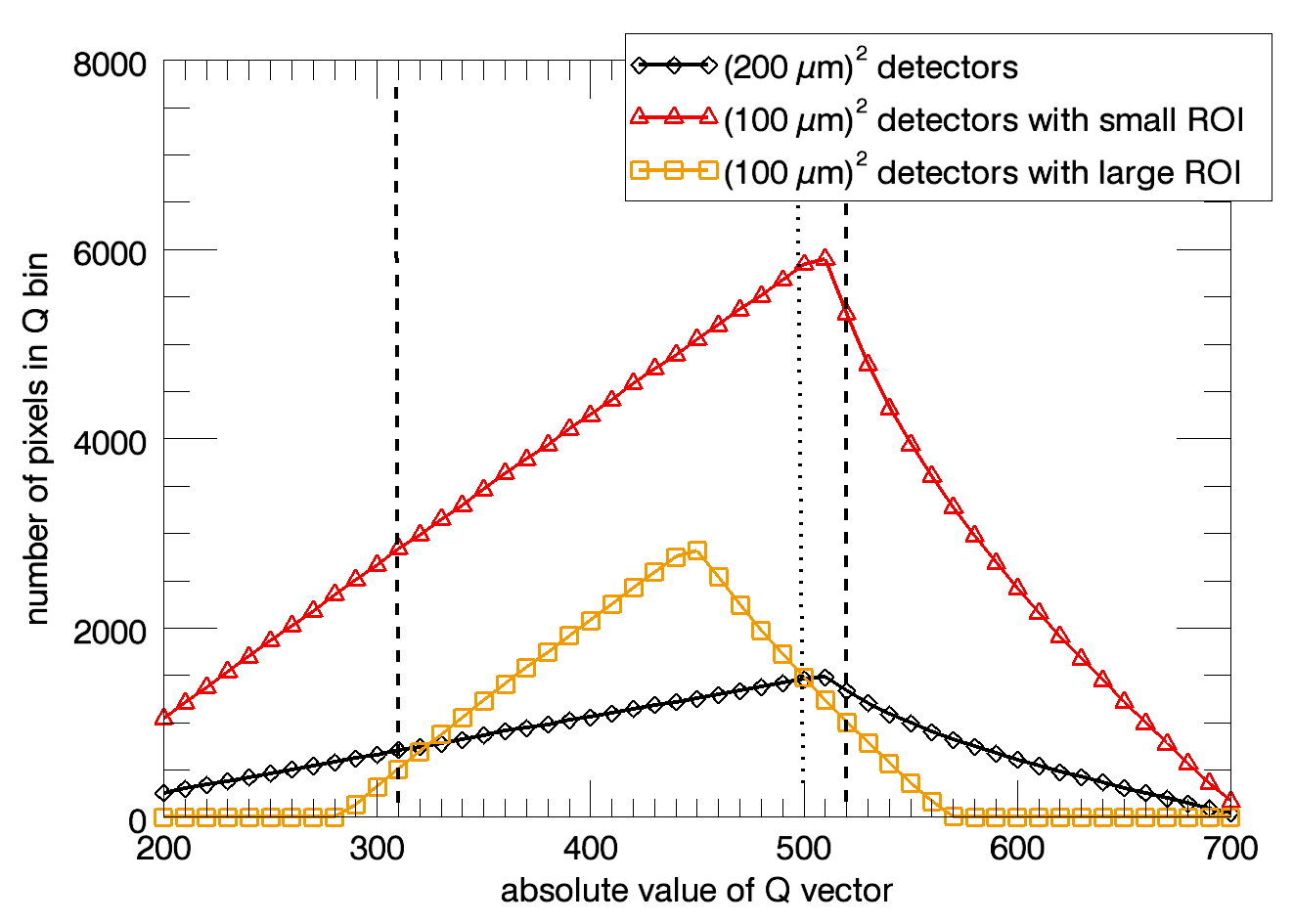}
	\caption{Number of pixels as a function of $Q$ value for the simulated detector systems. The $Q$ range between the dashed lines has been used for $Q$ averaging, as explained in the text. The dotted line is the $Q$ vector shown in Figures \ref{g2s100}, \ref{g2s200}, \ref{snr} and \ref{snr_zoom}.}
	\label{npix}
\end{figure}

Azimuthal averages (ensemble averages) were calculated by binning pixels that fulfilled the condition $Q-\Delta Q \leq Q(p) < Q+\Delta Q$, where $Q(p)$ was the absolute value of the $Q$ vector for pixel $p$ and $2\Delta Q=10$ was the bin size. 
Exemplary $Q$ space bins are shown in Figure \ref{areas}, the number of binned pixels as a function of $Q$ value is shown in Figure \ref{npix}. 
The ensemble averaged $g_2(Q,k\Delta t)$ function is the average of all $g_2(p,k\Delta t)$ functions in the corresponding $Q$ space bin.

As an example, the resulting $g_2(Q,k\Delta t)$ functions at a fixed $Q$ value of 500 units as a function of average intensity are shown for the ideal (100~$\upmu$m)$^2$ detector in Figure \ref{g2s100} and for AGIPD and MAAT in Figure \ref{g2s200}.

\begin{figure}[tb!]
	\centering
	\begin{tabular}{cc} 
		\begin{tabular}{c} \includegraphics[width=0.55\textwidth]{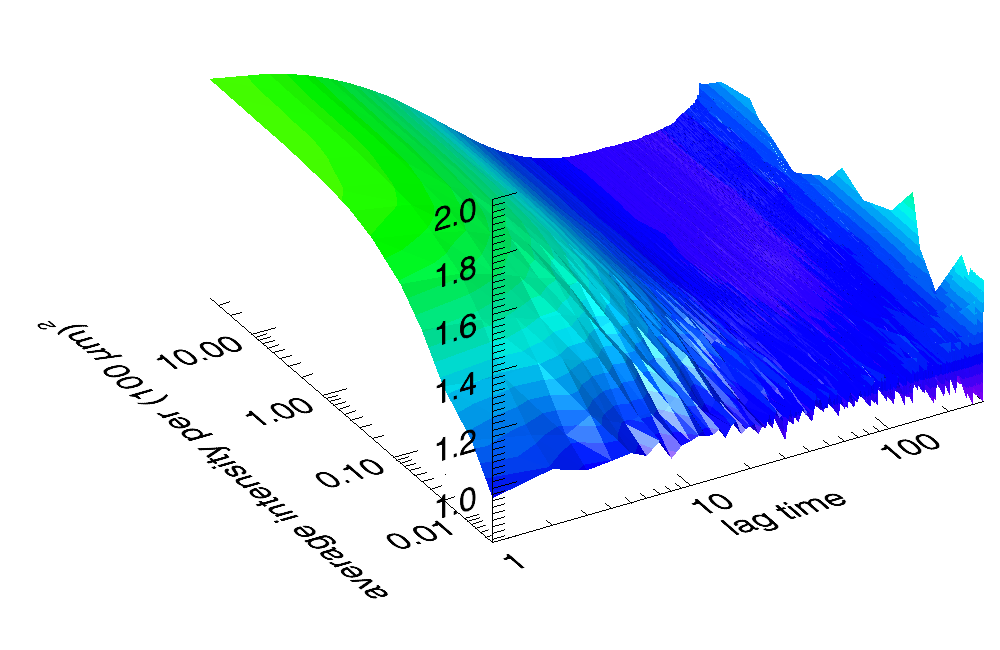} \end{tabular} &
		\begin{tabular}{c} \includegraphics[width=0.1\textwidth]{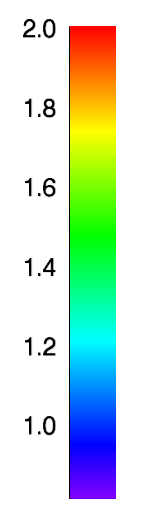} \end{tabular}
	\end{tabular}
	\caption{Resulting $g_2$ functions at a fixed $Q$ of 500 units as a function of average intensity. Results were obtained for the ideal (100~$\upmu$m)$^2$ detector for small ROIs. The results include photon statistics noise and noise from background photons and intensity fluctuations. The apparent increase of the $g_2$ at very large lag times is an artifact induced by the finite number of images. The noise is especially prominent at the lowest intensities.}
	\label{g2s100}
\end{figure}
	
Figure \ref{g2s100} shows the expected exponential decay of the autocorrelation function with increasing lag time. The maximum value for low lag times is approximately 1.7, thus below the theoretical maximum value of 2, as the experimental contrast is smaller than one. For lag times beyond 100 the function falls below 1.0 and increases again towards larger lag times. This behavior is commonly observed when evaluating the intensity autocorrelation function on a limited set of frames and would disappear if more frames were imaged in sequence. The apparent contrast decreases as the intensity decreases. At the lowest intensities the images become uncorrelated as they are dominated by noise, correspondingly the $g_2$ function appears flat.

\begin{figure}[tb!]
	\centering
	\begin{tabular}{cc} 
			\begin{tabular}{c} 
			  \includegraphics[width=0.55\textwidth]{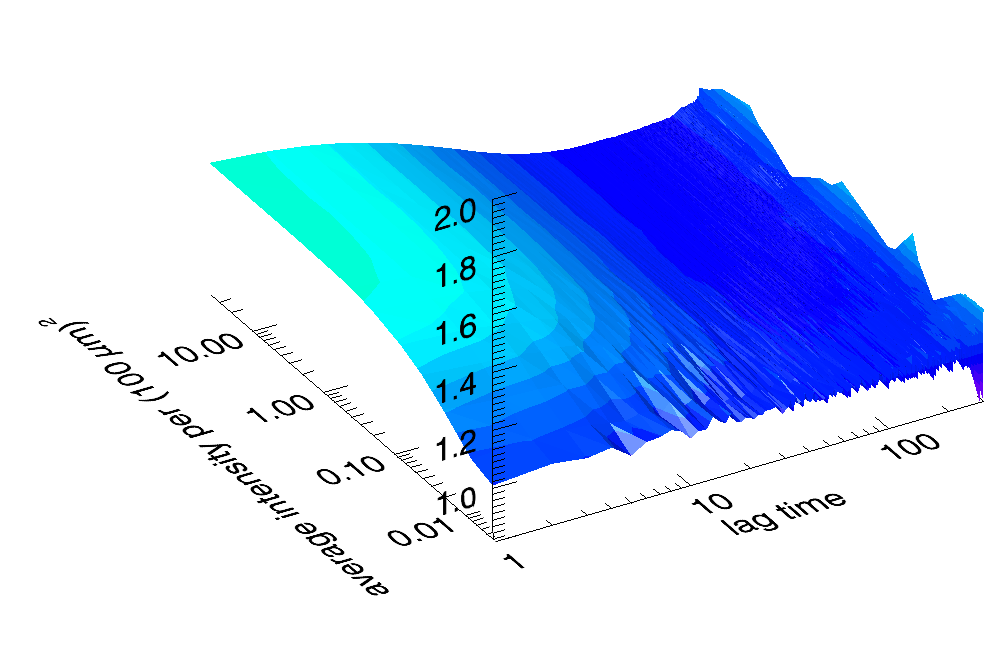} \\
			  \includegraphics[width=0.55\textwidth]{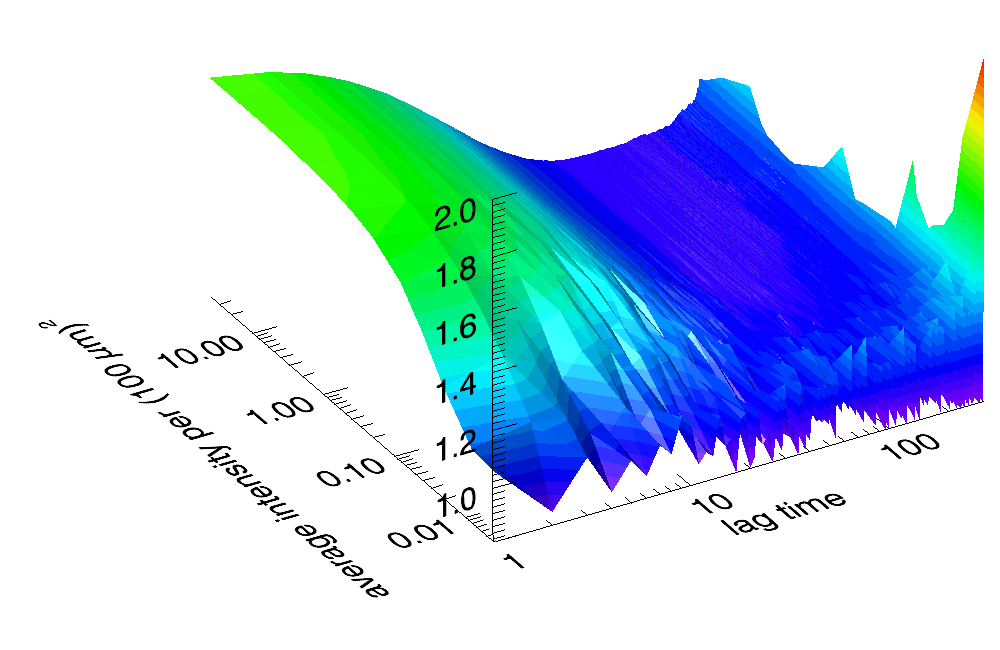} 
			\end{tabular}
		\begin{tabular}{c} \includegraphics[width=0.15\textwidth]{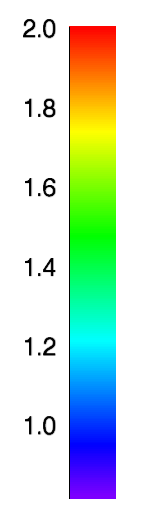} \end{tabular}
	\end{tabular}
	\caption{Resulting $g_2$ functions at a fixed $Q$ of 500 units as a function of average intensity. The upper image shows the results for the AGIPD, the lower image shows the results for the MAAT.  The results include photon statistics noise and noise from background photons and intensity fluctuations. The apparent increase of the $g_2$ at very large lag times is an artifact induced by the finite number of images. The noise is especially prominent at the lowest intensities.}
	\label{g2s200}
\end{figure}

Figure \ref{g2s200} shows the exponential decay of the autocorrelation function for the AGIPD and MAAT. The features seen for the ideal system are observed here as well. The maximum value for AGIPD and MAAT at low lag times is approximately 1.3 and 1.7, thus MAAT produces a similar contrast to the ideal (100~$\upmu$m)$^2$ detector. When compared to the $g_2$ functions shown in Figure \ref{g2s100}, the AGIPD curves appear smoother, as the larger pixel size averages over a larger area, while the MAAT curves appear less smooth, as the the $g_2$ function is calculated from less detected photons. This is especially prominent at the lowest intensities.

\subsection{Fit to the $g_2$ function}
\label{scale_par}

Each ensemble averaged $g_2(Q,k\Delta t)$ function was fitted with the following function to evaluate the performance of the different detector systems:

\begin{equation}
	g_2\left(Q,k\Delta t\right) = S(Q) \left\{ C(Q)\exp(-2\Gamma(Q) \tau) + 1\right\} \label{fit_funct}
\end{equation}
where $\tau=k\Delta t$ is the lag time, $S(Q)$ a scaling parameter, $C(Q)$ the optical contrast and $\Gamma(Q)$ the inverse of the correlation time $\tau_c$. 

For the dispersion relation $\Gamma(Q)$ it was assumed that $\Gamma(Q)=C_{disp} Q$, and likewise $\tau_c=\left( C_{disp} Q \right)^{-1}$. 

The scaling parameter $S(Q)$ was necessary to compensate the effect of pixels which detect zero photons during the bunch train. It can be calculated analytically as $1-\left( N_{\langle  I_p\rangle  =0}/N_Q \right)$, where $N_{\langle  I_p\rangle  =0}/N_Q$ is the fraction of pixels in a given $Q$ bin which have zero intensity in all frames of a given pulse train.

\section{Results}

The influence of the intensity fluctuation factor $R_{fluct}$ in Equation \ref{quant} on the results was found to be negligible. Thus two sets of results are compared here: the results with photon statistics noise alone  ($R_{fluct}=1.0$ and $F_{noise}=0$), and the results with photon statistics noise and noise from background photons and intensity fluctuations.

Using the Fourier transform to generate diffraction patterns implies an infinite coherence length, both longitudinal and transverse, thus the commonly observed drop in contrast at high $Q$ values is neglected. As a consequence there is little change in contrast and scaling factor as a function of $Q$ value. Thus the fit results were averaged for $Q$ values between 310 and 520 units. 

The $Q$ range between 310 and 520 units has been chosen as it excludes $Q$ vectors with low statistics towards the edge of the sensitive area for large regions of interest. The (200~$\upmu$m)$^2$ pixel detectors and small ROI cases have been restricted to the same $Q$ range to have a comparable data set, although the errors for these systems could have been reduced using the additional information available at other $Q$ values.

\subsection{Scaling parameter}

\begin{figure}[tb!]
	\centering
	$\begin{array}{cc} 
		\includegraphics[width=0.5\textwidth]{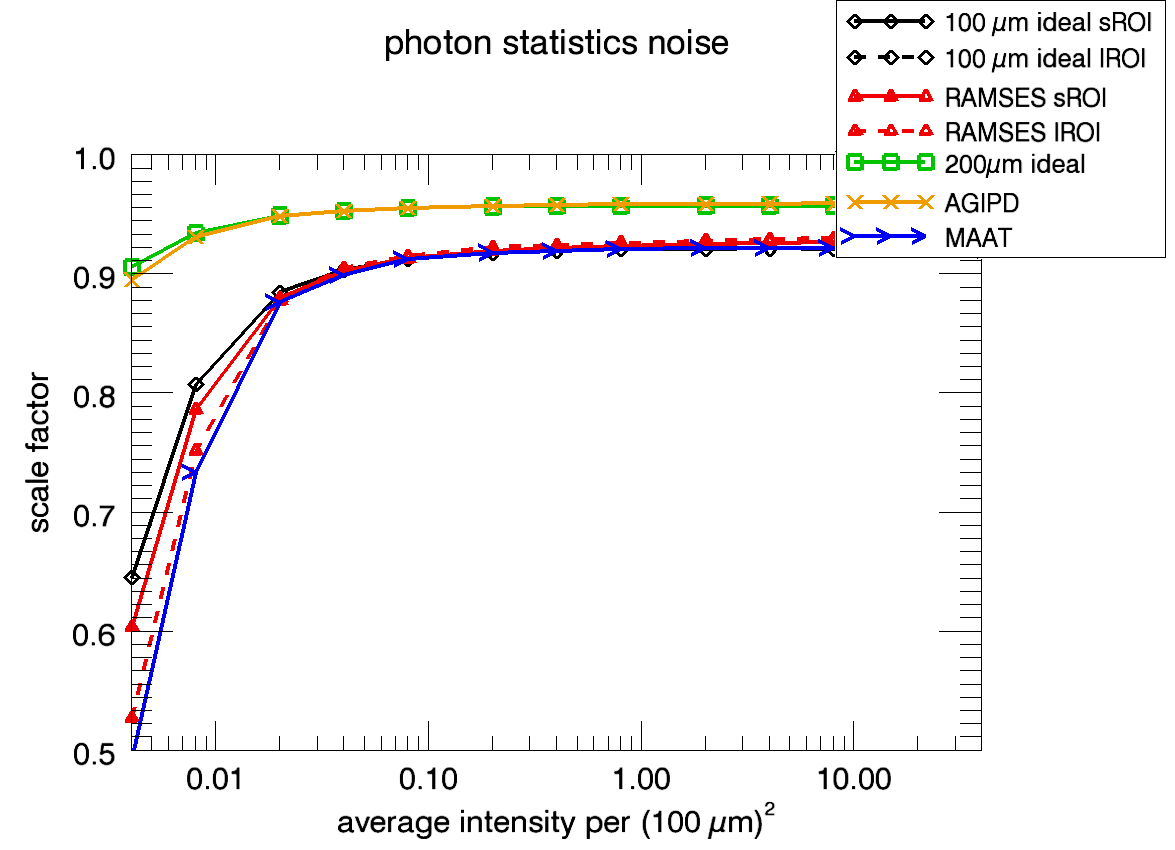} &  \includegraphics[width=0.5\textwidth]{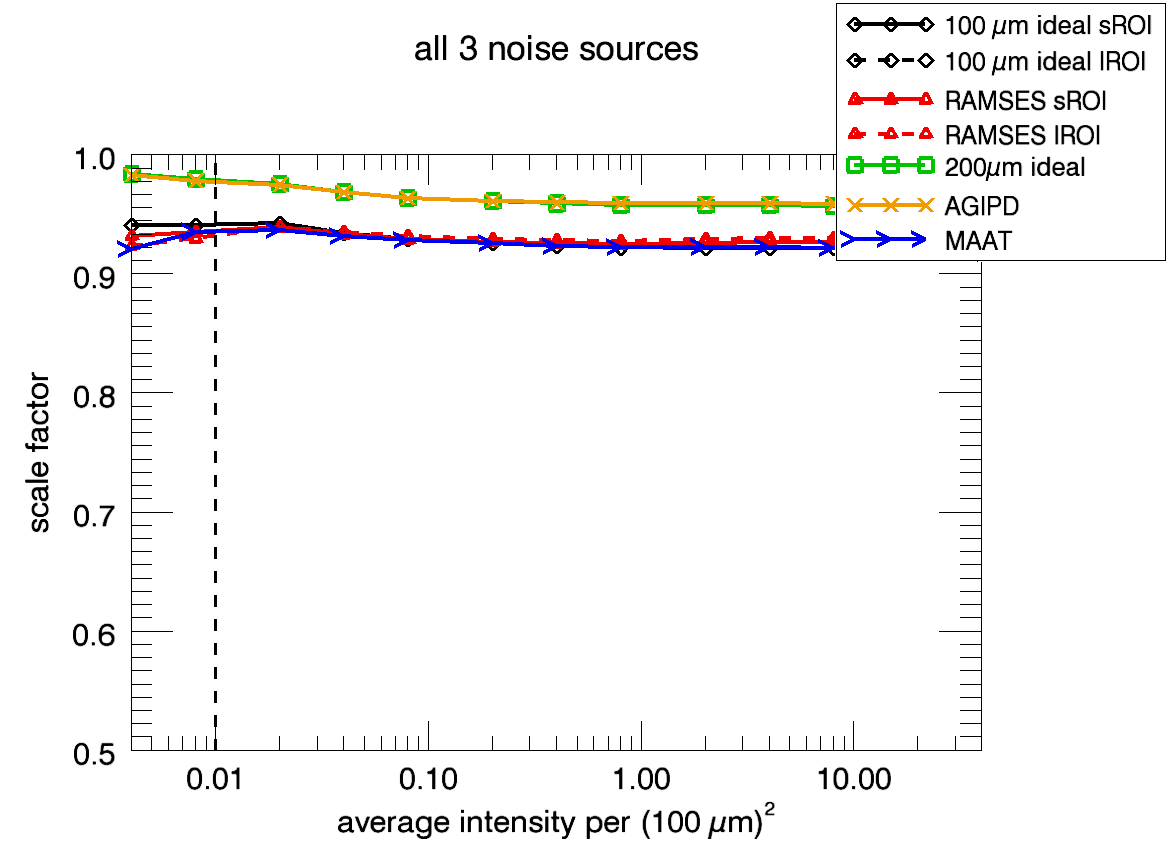} \\
	\end{array}$
	\caption{Average scaling parameter for all systems. The left shows the results for photon statistics noise only, the right image shows the results including photon statistics noise and noise from background photons and intensity fluctuations. The dashed vertical line indicates the average intensity of $F_{noise}$.}
	\label{scale}
\end{figure}

Generally the scaling parameter shown in Figure \ref{scale} behaves as expected and explained in section \ref{scale_par}. The average scaling parameter for the (200~$\upmu$m)$^2$ systems is always larger than for the (100~$\upmu$m)$^2$ system as the probability to measure photons increases with the sensitive area.

When summing together all the images of a pulse train, the scaling factor can be interpreted as the fraction of non-zero pixels in the summed image. This can be clearly observed as without noise, the scaling factor is decreasing at low intensities. 

For the summed image the expected number of noise photons $N_{\gamma,noise}$ can be analytically calculated to be $N_{\gamma,noise}=N_F/P_{noise}=3.0$ photons per (100~$\upmu$m)$^2$. Since these photons follow Possionian statistics, the probability to have zero photons in a pixel of the summed image when the expected value is 3 is $P_3(0) \approx 0.05$. When noise is present an average scaling factor of $1-P_3(0) \approx 0.95$ is expected for (100~$\upmu$m)$^2$ systems, and this is observed for low intensities.

\subsection{Optical contrast}

\begin{figure}[tb!]
	\centering
	$\begin{array}{cc} 
		\includegraphics[width=0.5\textwidth]{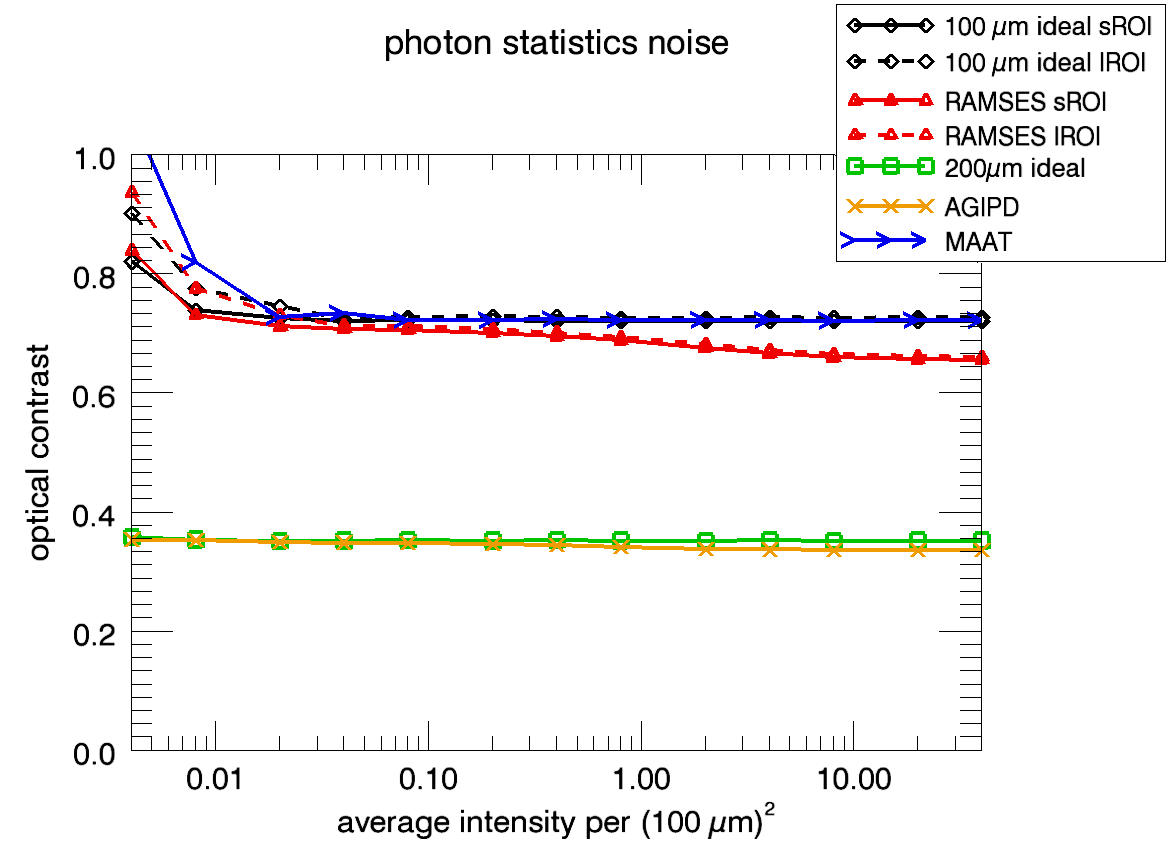} &  \includegraphics[width=0.5\textwidth]{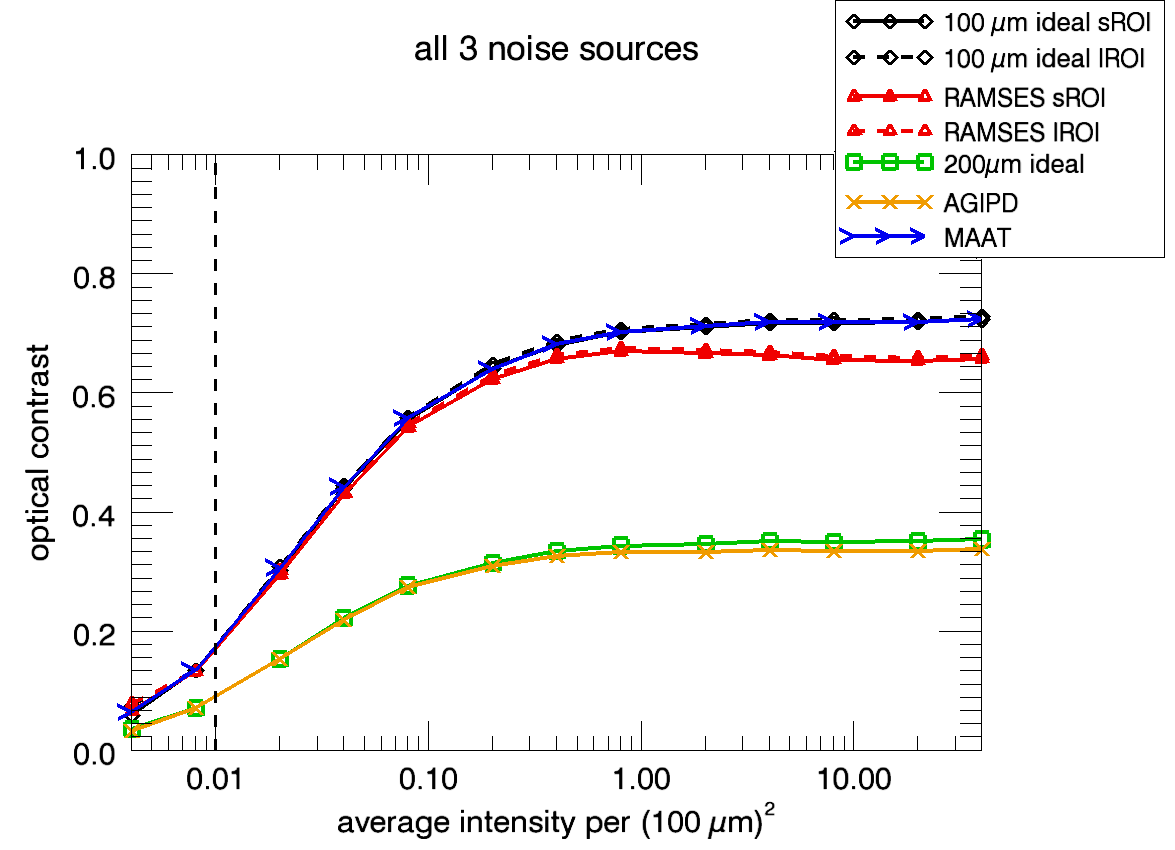} \\
	\end{array}$
	\caption{Average optical contrast for all detector systems as a function of intensity. The left image shows the results for photon statistics noise only, the right image shows the results including photon statistics noise and noise from background photons and intensity fluctuations. The dashed vertical line indicates the average intensity of $F_{noise}$.}
	\label{contrast}
\end{figure}

Figure \ref{contrast} shows the $Q$ averaged results for the optical contrast. When $F_{noise}$ is included in the simulations the optical contrast decreases once $\langle  I\rangle  /I_{noise} \lesssim 100$ ($\langle  I\rangle   \approx 1$), while it stays almost constant without noise. For $F_{noise}=0$ and very low intensities the contrast apparently increases due to the biasing effect of the scaling factor. However this additional contrast did not improve the data quality, as the scaling factor decreased strongly.

Figure \ref{contrast} shows that, except for MAAT, the real systems show a lower contrast than the ideal ones, independent of the noise. The effect is small for the AGIPD, as the vast majority of photons do not undergo charge sharing, but the observed small difference between the green and yellow lines in Figure \ref{contrast} is statistically significant. At low intensities the difference vanishes. This reduced contrast was caused by charge sharing events, which correlate the pixel values. Charge sharing is excluded for MAAT by its design. 

The fact that the results for MAAT reproduce the results for an ideal (100~$\upmu$m)$^2$ system, indicates that the detector noise and quantum efficiency do not influence the data quality.

For the ideal systems the contrast was given by $C(\langle  I\rangle  ) \approx C_{geom} (1-I_{noise}/\langle  I\rangle  )$, where $C$ is the determined contrast and $C_{geom}$ the contrast expected from geometric estimations:

\begin{equation}
	C_{geom} = \frac{1}{1+\frac{P^2}{S^2}} \label{c_geo}
\end{equation}
with P being the linear pixel size and S the full width at half maximum of the speckles.

The geometric estimate and the results of the ideal systems at the highest intensity are in good agreement (0.68 and 0.72 or 0.35 and 0.36 for (100~$\upmu$m)$^2$ or (200~$\upmu$m)$^2$ systems, respectively). 

\subsection{Correlation time}

\begin{figure}[tb!]
	\centering
 	\includegraphics[width=0.55\textwidth]{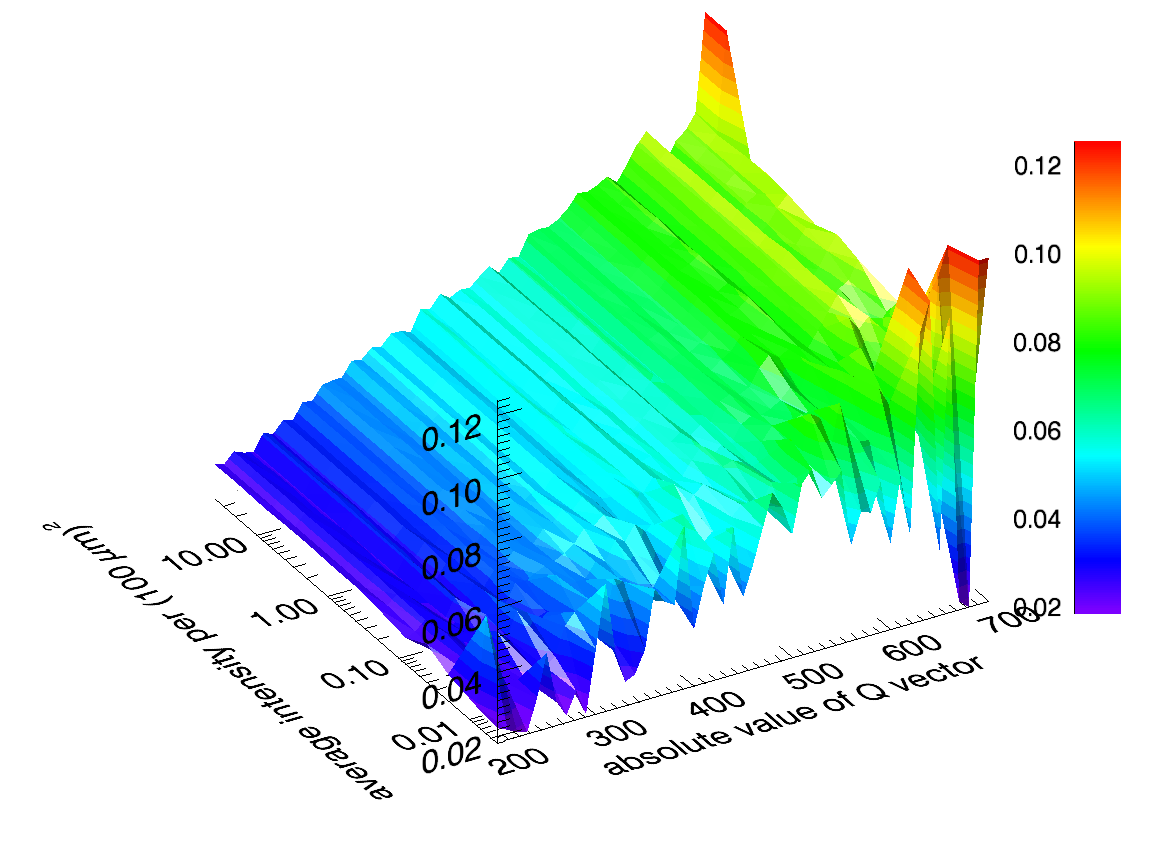} 
	\caption{Inverse correlation time $\Gamma(Q)$ for the ideal (100~$\upmu$m)$^2$ detector and small ROIs as a function of $Q$ vector and intensity for photon statistics noise only. The results for the other detector systems look similar, but show larger variations.}
	\label{corr1}
\end{figure}

Figure \ref{corr1} shows the inverse correlation time $\Gamma(Q)$ for the ideal (100~$\upmu$m)$^2$ pixel detector and small ROIs as a function of $Q$ vector and intensity.

Overall, the correlation time was observed to be approximately inversely proportional to the $Q$ vector ($\Gamma(Q)=C_{disp} Q$, likewise $\tau_c=\left( C_{disp} Q \right)^{-1}$). The $Q$ averaged results for the dispersion constant $C_{disp}$ are displayed in Figure \ref{corr2} for all simulated detector systems.

For purely diffusive systems $\Gamma(Q)$ should be proportional to $Q^2$, but the investigated jump diffusion system did not show such a behavior. Literature shows \cite{jumpdiff1, jumpdiff2} that a diffusive behavior is only expected for low $Q$ vectors, where the length scale is large compared to the jump size. 

As the low $Q$ vectors are not included in the data evaluation (detailed in \ref{image}), a $Q^2$ dependence is neither observed nor expected.

\begin{figure}[tb!]
	\centering
	$\begin{array}{cc} 
		\includegraphics[width=0.5\textwidth]{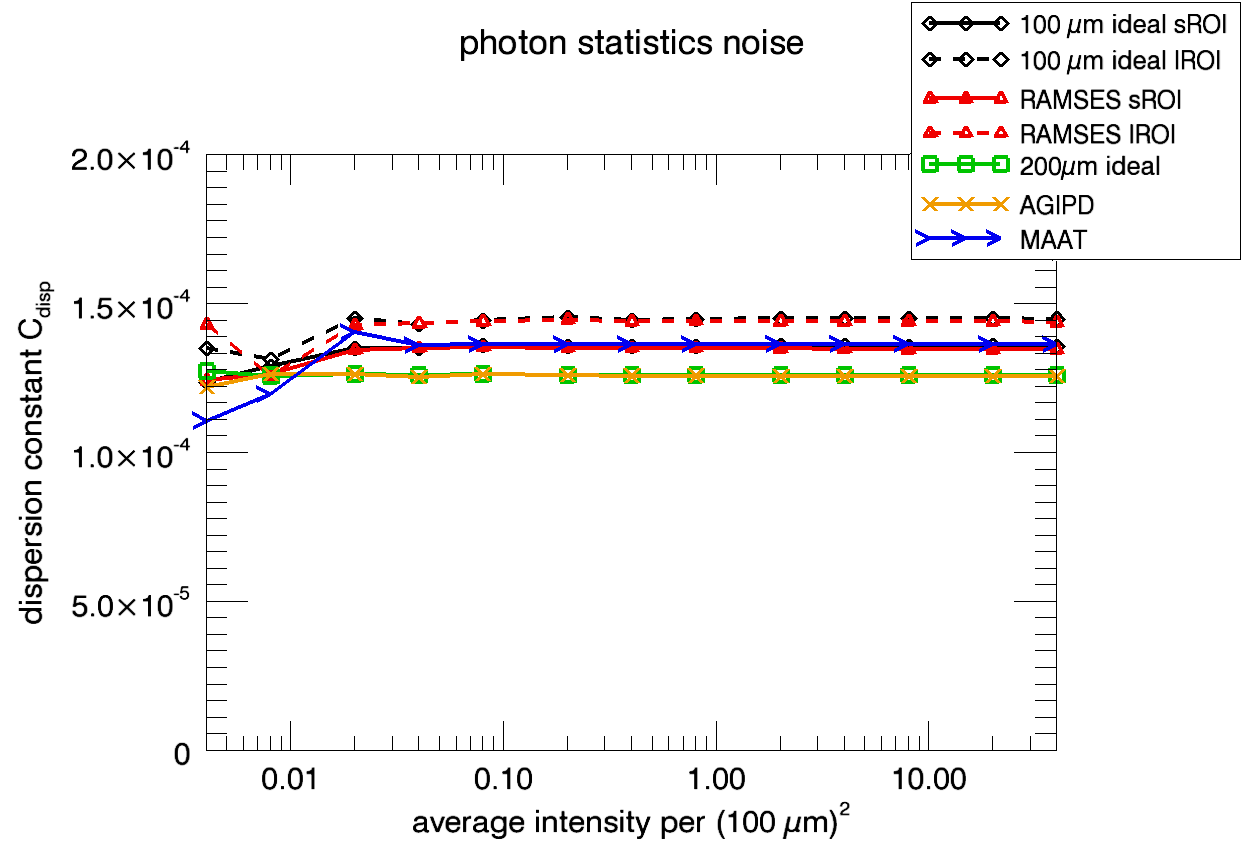} & \includegraphics[width=0.5\textwidth]{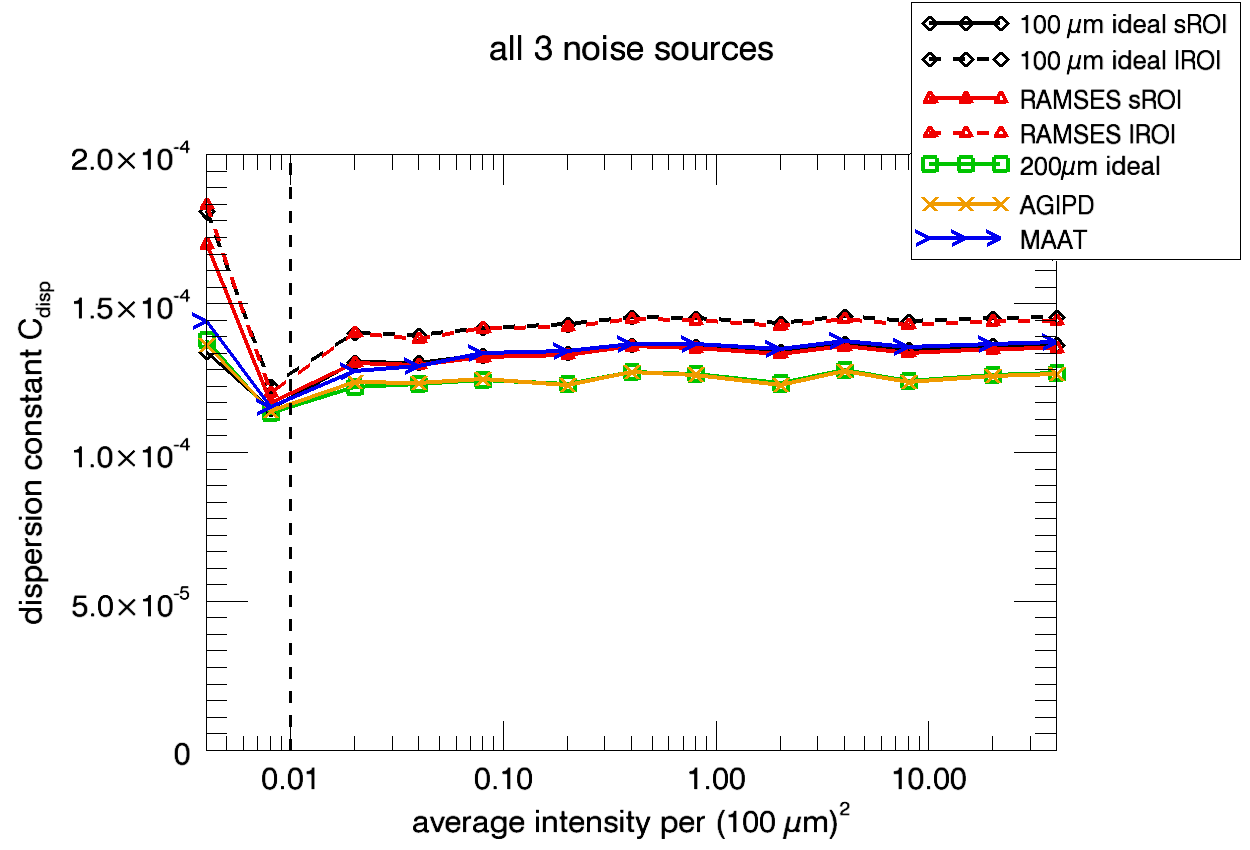} \\
	\end{array}$
	\caption{Dispersion constant $C_{disp}$ for all detector systems as a function of average intensity. The left image shows the results for photon statistics noise only, the right image shows the results including photon statistics noise and noise from background photons and intensity fluctuations. The dashed vertical line indicates the average intensity of $F_{noise}$.}
	\label{corr2}
\end{figure}

Figure \ref{corr2} shows a slightly different dispersion constant $C_{disp}$ for each system. This is due to the different $Q$ space coverage of the systems and different averaging properties of the different pixel sizes and the crude approximation of an inversely proportional dispersion relation, which is purely empiric. However the stability of the results as a function of intensity show that this is merely a systematic error, which is identical for all systems of the same type, so it will not influence possible conclusions.

When evaluating the influence of the average intensity in Figure \ref{corr2} one notices instabilities for the results at average intensities $\lesssim$ 0.01. At these intensities the fit results are of low statistical significance. The extracted dispersion constant $C_{disp}$ was observed to be more sensitive to the low significance than the contrast. A reason for this might be the additional data evaluation step necessary to extract $C_{disp}$.

When evaluating the influence of the added incoherent noise by comparing the left to the right image in Figure \ref{corr2} at average intensities $\gtrsim$ 0.02, one notices very little effect. The reason of this is that the correlation time is independent of the contrast, as long as the contrast is high enough to reliably determine the correlation time.

\section{Signal to noise ratio}

\begin{figure}[tb!]
	\centering
		\includegraphics[width=0.8\textwidth]{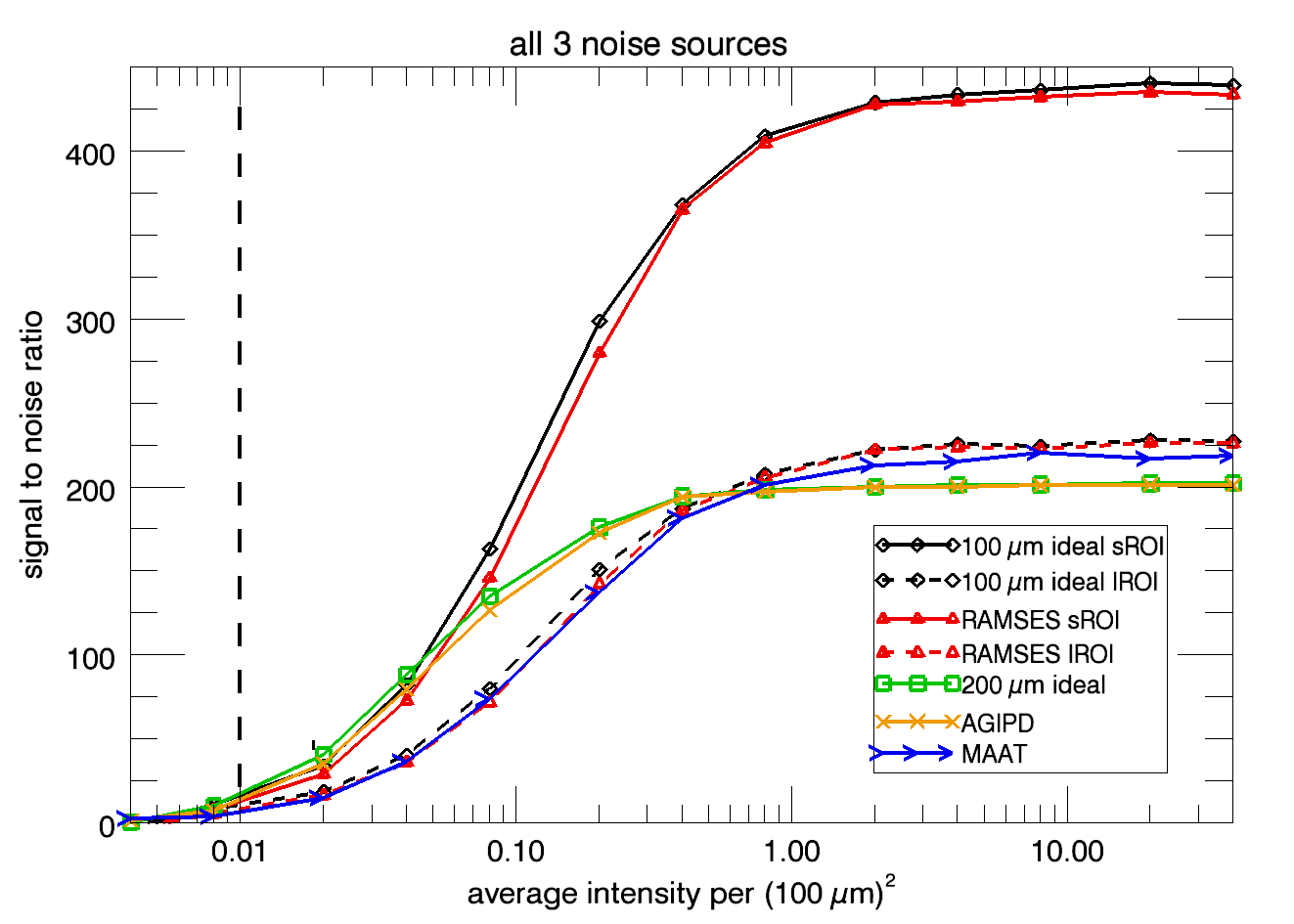} 
	\caption{Signal to noise ratio for all detector systems as a function of average intensity including photon statistics noise and noise from background photons and intensity fluctuations. The curves shown have been determined for a $Q$ value of 500, other $Q$ values produce similar results. The dashed vertical line indicates the average intensity of $F_{noise}$.}
	\label{snr}
\end{figure}

The signal to noise ratio is commonly used as a figure of merit for XPCS experiments. Adapted to the situation of this study the analytic expression for the SNR, derived from the mean value of the $g_2$ function and its error ($err(g_2)$)\footnote{The error of the mean $\bar{X}$ using $N_s$ samples of a statistical population $X$ is usually defined as $err(\bar{X})=\sqrt{var_s(X)/N_s}$, where $var_s(X)=\frac{1}{N_s-1}\sum_i{(X_i-\bar{X})^2}$ is the variance of the $N_s$ samples of $X$.}, is:

\begin{equation}
	SNR(\langle  I\rangle  ) = \frac{g_2 -1}{err(g_2)} \propto \langle  I_p\rangle  C(\langle  I\rangle  )\sqrt{N} \label{snr_eq}
\end{equation}
with $\langle  I_p\rangle  $ being the average intensity in a pixel area, $C(\langle  I\rangle  )$ the optical contrast at this intensity and N the number of detector pixels.

Figure \ref{snr} shows the signal to noise ratio determined by dividing the first value ($k \Delta t=1$) of the $g_2$ function minus 1 by its error determined from the variance of all $g_2$ functions at this lag time in the same Q space bin. This is the definition of the signal to noise ratio in the limit of small lag times as indicated in Equation \ref{snr_eq}. Figure \ref{snr} shows the signal to noise ratio at a fixed Q value of 500 (indicated by the dotted line in Figure \ref{npix}). Evaluation of the signal to noise ratio at different Q values showed identical results.

The main features of Figure \ref{snr} are the increase in SNR with intensity for low intensities and the saturation of the SNR for high intensities. The saturation value is reached around average intensities of 1 photon per (100~$\upmu$m)$^2$. Additionally Figure \ref{snr} shows multiple separate sets of lines, one set for (100~$\upmu$m)$^2$ with small ROI, one set for large ROI and one set for (200~$\upmu$m)$^2$ systems. For low intensities the signal to noise ratio of the (200~$\upmu$m)$^2$ systems is approximately equal to the SNR of (100~$\upmu$m)$^2$ systems with small ROI and approximately twice the SNR of the (100~$\upmu$m)$^2$ systems with large ROI. Towards saturation all (100~$\upmu$m)$^2$ systems show larger SNRs than the (200~$\upmu$m)$^2$ systems, although only a little for large ROI and MAAT. The ratio of the SNR between large and small ROIs corresponds the square root of the ratio of the number of pixels. Ideal systems produce an SNR which is a little higher than that of the real systems due to their perfect quantum efficiency.

\begin{figure}[tb!]
	\centering
	$\begin{array}{cc} 
		\includegraphics[width=0.5\textwidth]{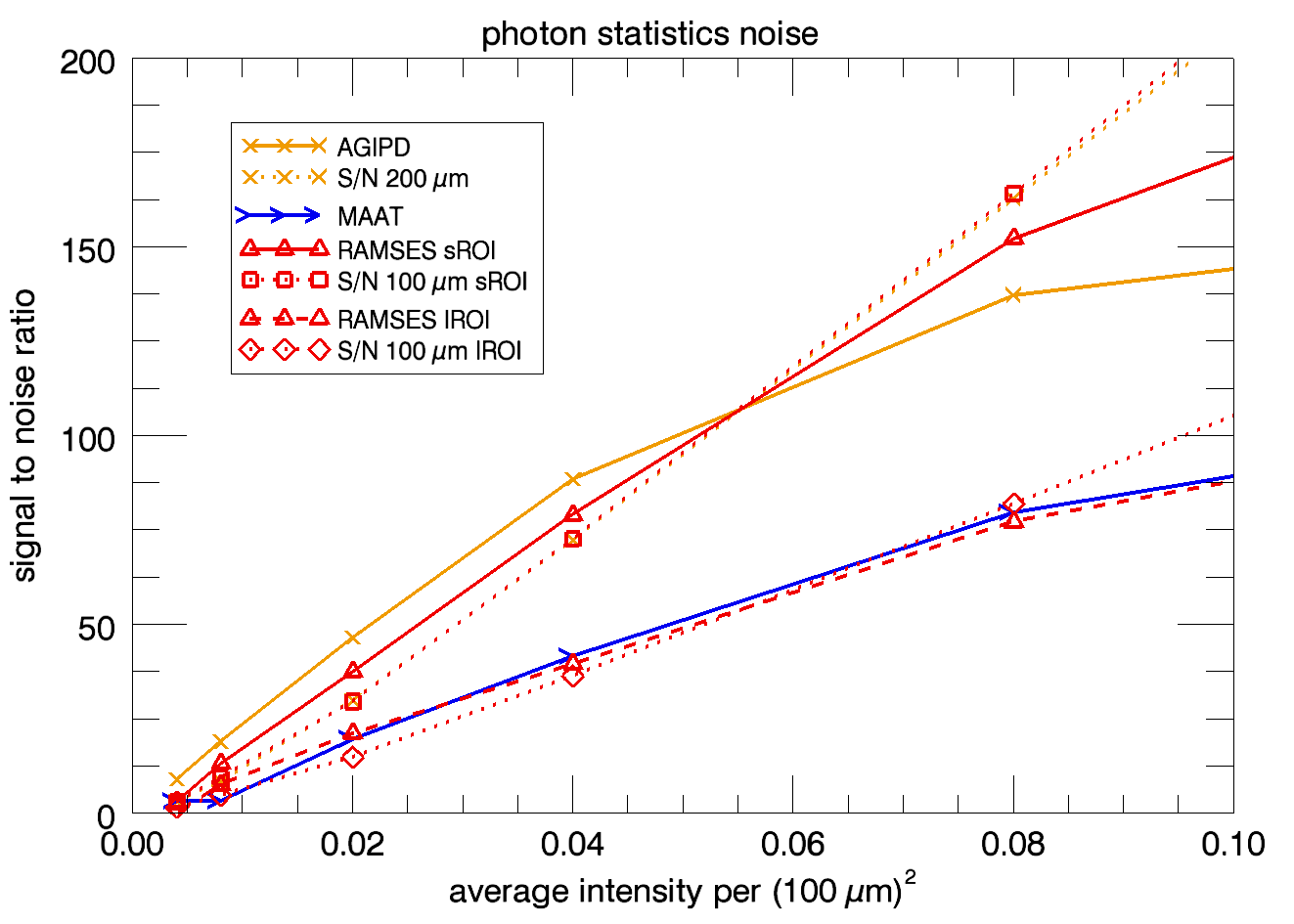} &  \includegraphics[width=0.5\textwidth]{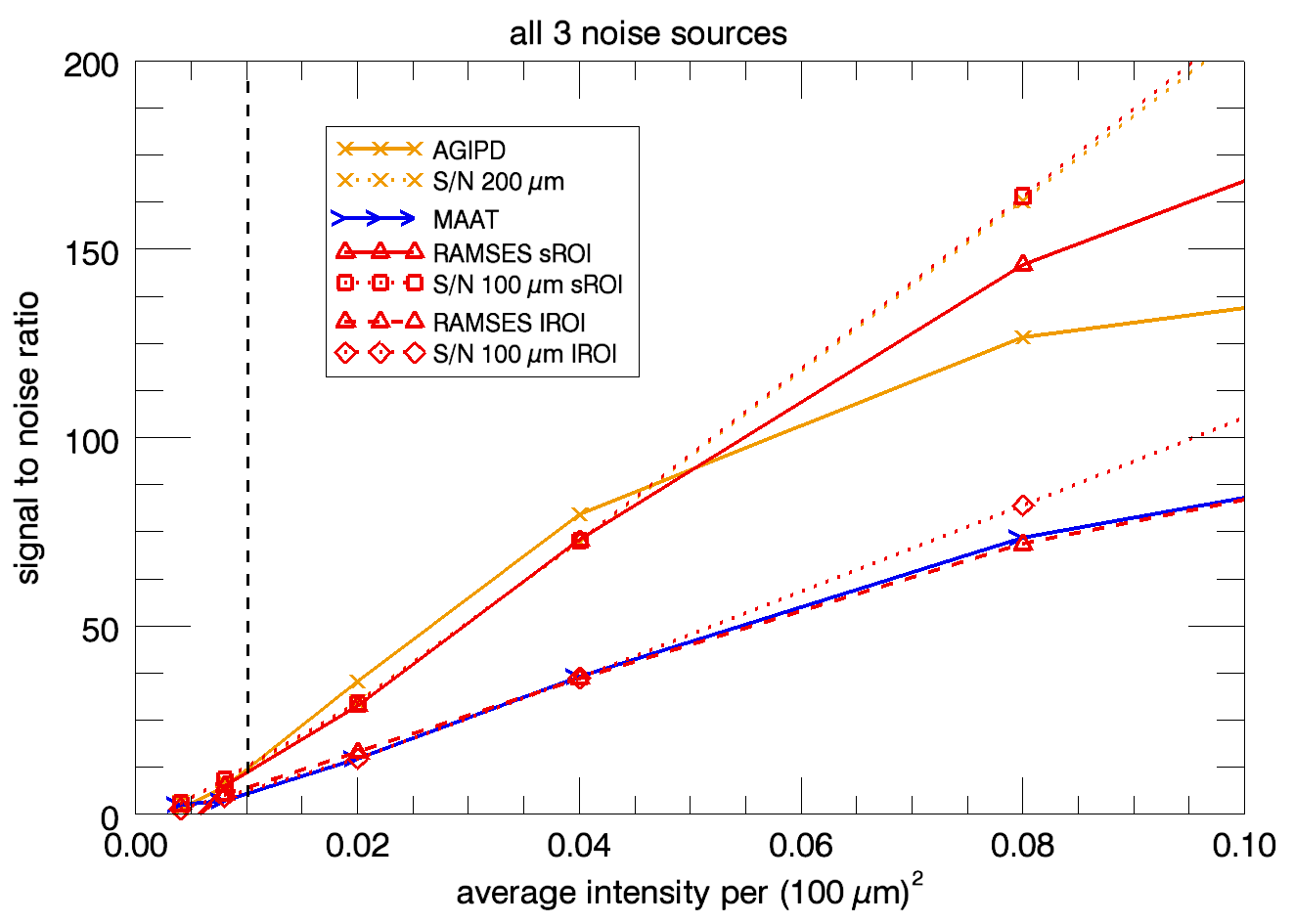} \\
	\end{array}$
	\caption{Signal to noise ratio for selected detector systems in the low intensity regime. The left image shows the results for photon statistics noise only, the right image shows the results including photon statistics noise and noise from background photons and intensity fluctuations. The dashed vertical line indicates the average intensity of $F_{noise}$. The analytical SNR for (200~$\upmu$m)$^2$ systems is almost identical to the analytical SNR for (100~$\upmu$m)$^2$ systems and small regions of interest; their lines coincide almost completely.}
	\label{snr_zoom}
\end{figure}

Equation \ref{snr_eq} was derived under the assumption of low count rates and small contrast in the limit of small lag times (\cite{jakeman, falus} and references therein). A more elaborate derivation, also for large lag times, can be found in \cite{schaetzel}. According to Equation \ref{snr_eq} the SNR increases with intensity until the assumptions for its derivation are no longer valid.

Using Equation \ref{snr_eq} to compare the SNRs of two systems with linear pixel sizes $P_1$ and $P_2$ yields:

\begin{equation}
	\frac{SNR_1(\langle  I\rangle  )}{SNR_2(\langle  I\rangle  )} = \frac{\langle  I_1\rangle  }{\langle  I_2\rangle  }\frac{C_1(\langle  I\rangle  )}{C_2(\langle  I\rangle  )}\sqrt{\frac{N_1}{N_2}} \label{snr_eq2}
\end{equation}

Assuming that the intensity in the pixel is proportional to its area and the intensity dependent terms of the contrast cancel, one can rewrite Equation \ref{snr_eq2} using Equation \ref{c_geo} as:

\begin{equation}
	\frac{SNR_1}{SNR_2} = \frac{P_1^2}{P_2^2}\frac{S^2+P_2^2}{S^2+P_1^2}\sqrt{\frac{N_1}{N_2}} \label{snr_eq3}
\end{equation}

Equation \ref{snr_eq3} yields values from $\sqrt{N_1/N_2}$ for $P_1,P_2 \gg S$ to $(P_1/P_2)^2 \sqrt{N_1/N_2}$ for $P_1,P_2 \ll S$. For the systems investigated here Equation \ref{snr_eq3} yields the results observed for low intensities, namely \newline $SNR_{100 \mu m, sROI} / SNR_{200 \mu m} \approx 1.0$, $SNR_{100 \mu m, lROI} / SNR_{200 \mu m}\approx 0.5 $ and \newline $SNR_{100 \mu m, sROI} / SNR_{100 \mu m, lROI} \approx 2.0$

Figure \ref{snr_zoom} shows the results for selected detector systems in the low intensity regime compared to the results from Equation \ref{snr_eq}. It can be seen that Equation \ref{snr_eq} is a well suited approximation of the SNR at low intensities. Furthermore it can be seen, that for intensities below approximately 0.05 photons per (100~$\upmu$m)$^2$ the signal to noise ratio of the AGIPD is larger than estimated from Equation \ref{snr_eq}.

\section{Relative error $E$ of the dispersion constant $C_{disp}$}

\begin{figure}[tb!]
	\centering
	$\begin{array}{cc} 
		\includegraphics[width=0.5\textwidth]{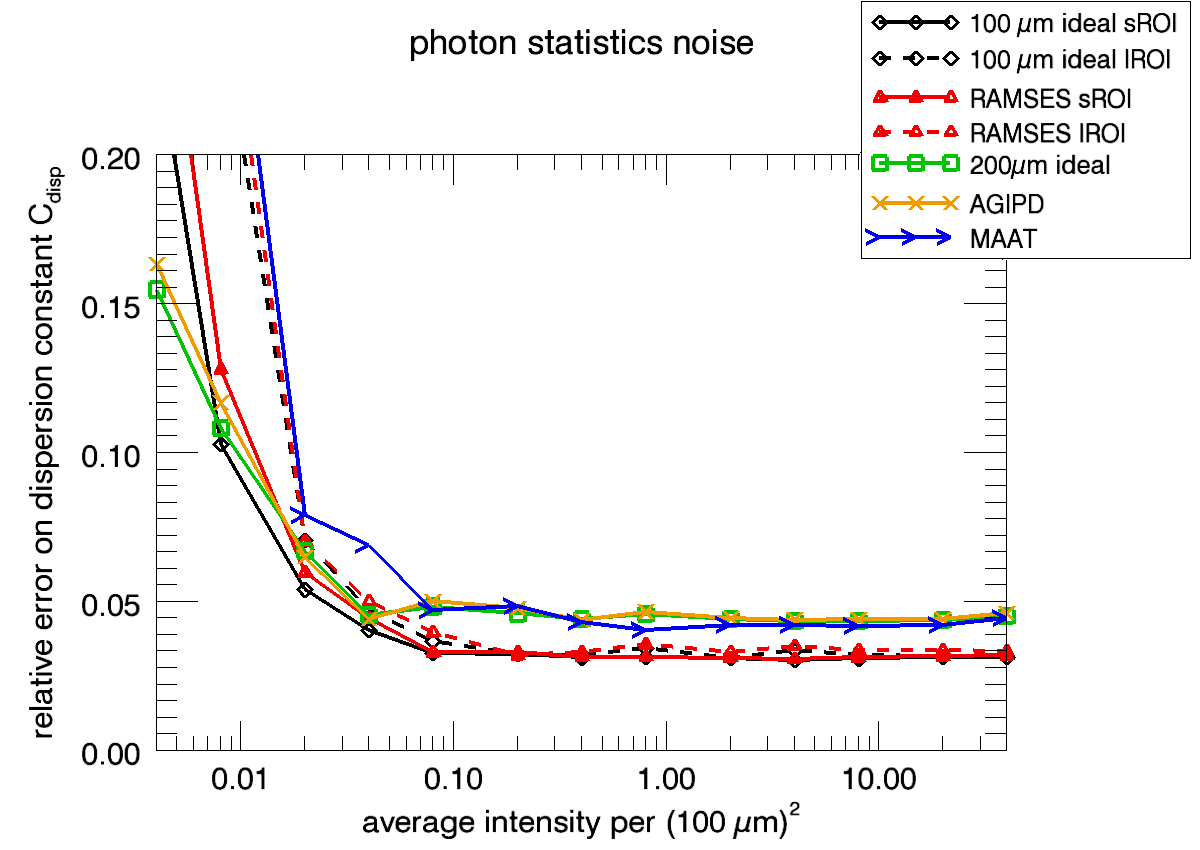} & \includegraphics[width=0.5\textwidth]{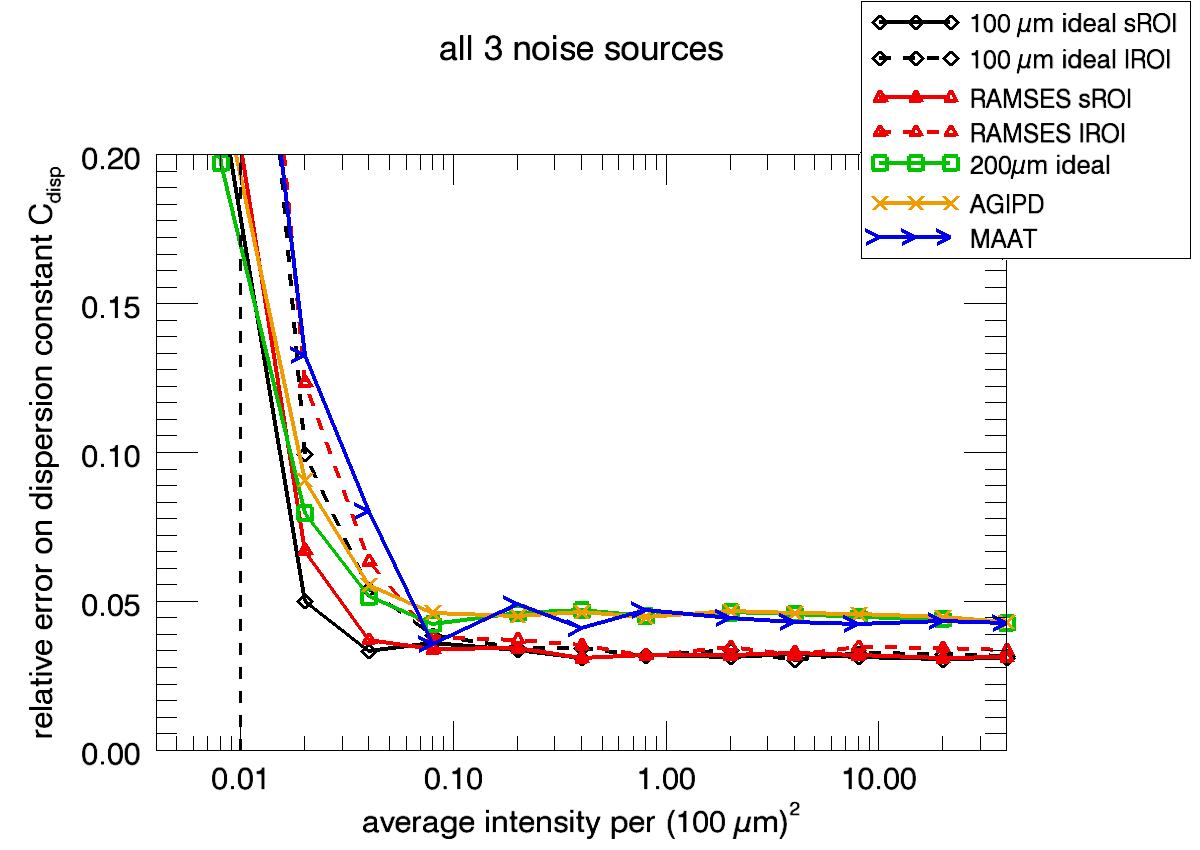} \\
	\end{array}$
	\caption{Relative error $E$ of the dispersion constant $C_{disp}$ for all detector systems as a function of average intensity. The left image shows the results for photon statistics noise only, the right image shows the results including photon statistics noise and noise from background photons and intensity fluctuations. The dashed vertical line indicates the average intensity of $F_{noise}$.}
	\label{error}
\end{figure}

\begin{figure}[tb!]
	\centering
	$\begin{array}{cc} 
		\includegraphics[width=0.5\textwidth]{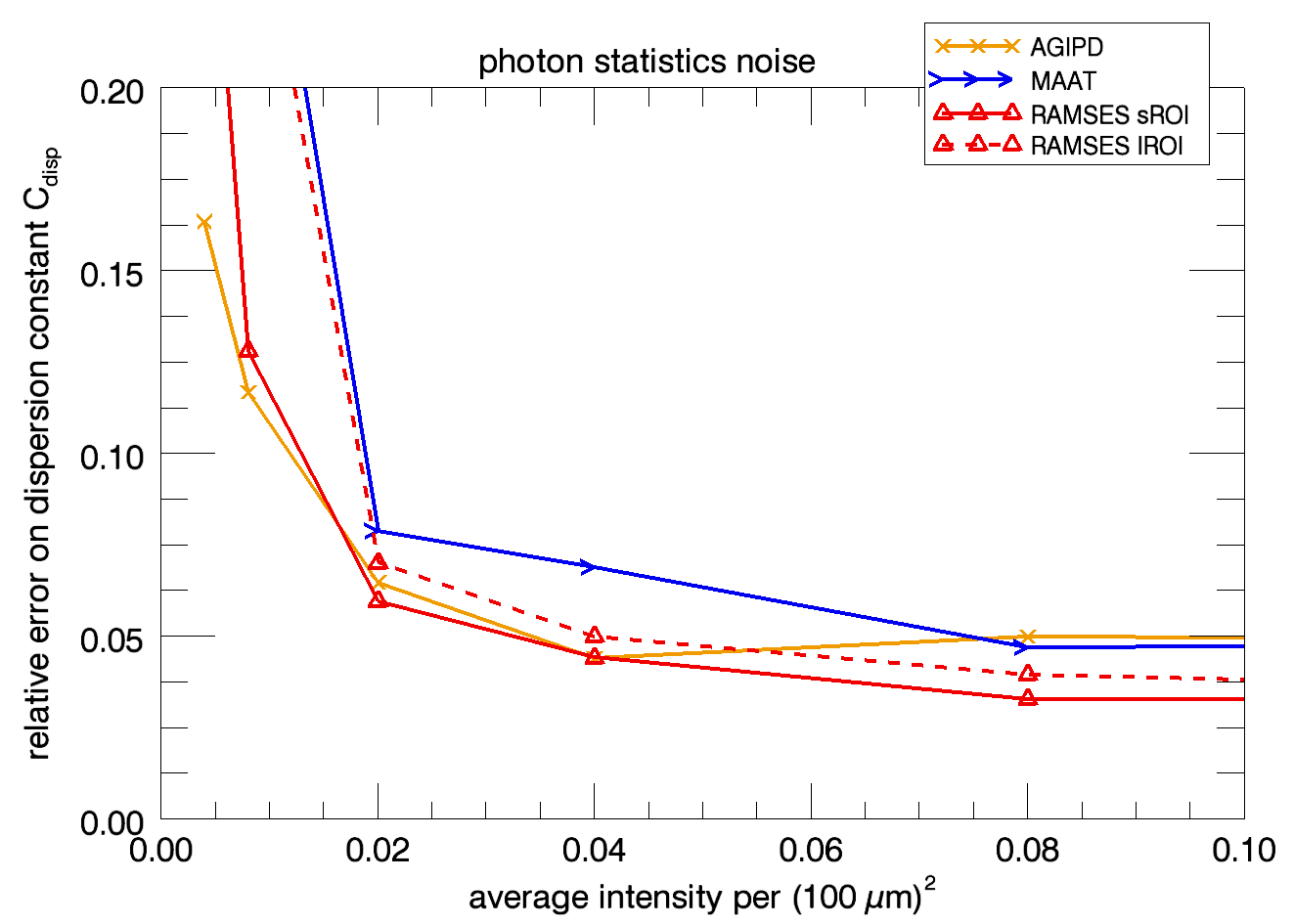} &  \includegraphics[width=0.5\textwidth]{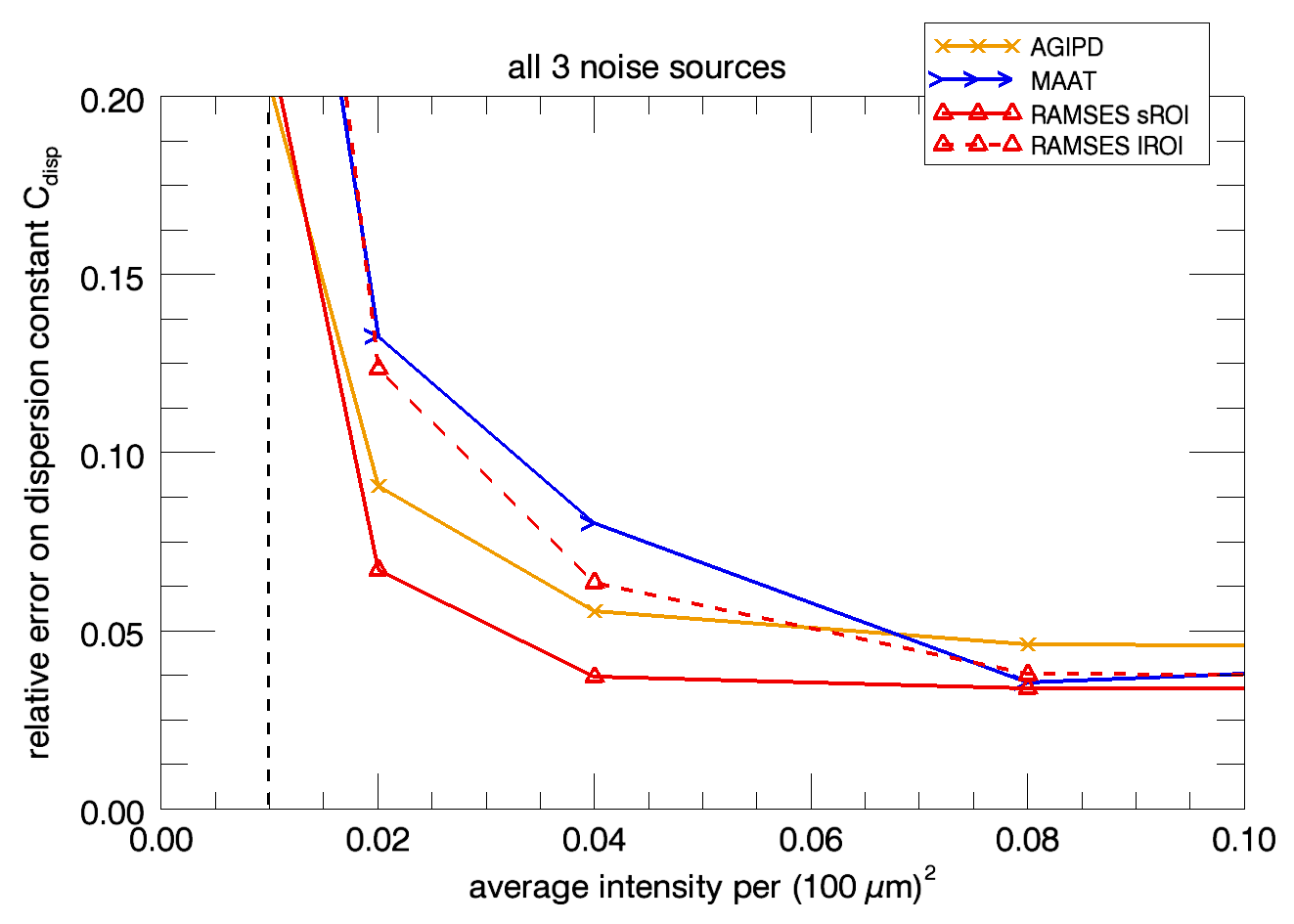} \\
	\end{array}$
	\caption{Relative error $E$ of the dispersion constant $C_{disp}$ for selected detector systems in the low intensity regime. The left image shows the results for photon statistics noise only, the right image shows the results including photon statistics noise and noise from background photons and intensity fluctuations. The dashed vertical line indicates the average intensity of $F_{noise}$.}
	\label{zoom}
\end{figure}

The focus here is on a slightly different aspect of the error determination in XPCS experiments. As the main physical quantity of interest are the time constants connected with the dynamics the relative error in the time constants given by the ratio $E = \sqrt{var(C_{disp})} / C_{disp}$ was investigated. This quantity was derived by fitting the $g_2$ functions with the exponential correlation functions (Eq. \ref{fit_funct}). Thus $E$ represents a different figure of merit than calculated by Equation \ref{snr_eq}. While for the SNR higher values indicate better quality, lower values of $E$ indicate better values.

Both images of Figure \ref{error} show a similar behavior in dropping from very large values at low intensities to a saturation value at high intensities. As already observed for the correlation time, the errors are very large for intensities $\lesssim$ 0.01, indicating problems with the proper determination of the resulting quantities. 

Comparing the left to the right image reveals the influence of the additional noise. While the data without additional noise is still usable at very low intensities the error increases significantly when the average intensity of the signal becomes comparable to the average noise intensity.

For average intensities above approximately 0.1 the results are independent of the noise.
From this point on the error contributions of the photon statistics and all other noise sources are small compared to the systematic errors. The systematic errors are introduced by artifacts in the $g_2$ functions causing a small, but systematic deviation of the $g_2$ functions from the function used for the fit. The artifacts in the $g_2$ functions are introduced as the correlation time is a significant fraction of the total sampling time, which is limited by the total number of frames\footnote{A possible remedy for this might be the use of alternate (e.g. logarithmic) sampling schemes, which will be investigated in a future study.}.

The saturation values are qualitatively different from the saturation values of the signal to noise ratio. The relative error saturates at lower intensities than the signal to noise ratio and to qualitatively different values, which result from the different numbers of pixels for the different detector systems, but do not show a $\sqrt{N}$ behavior. The evaluated data is restricted to the $Q$ range of 310 - 520 (shown in Figure \ref{npix}), so the results were biased in favor of the (100~$\upmu$m)$^2$ detector systems, as more pixels were used for the evaluation than for the (200~$\upmu$m)$^2$ detectors.

Figure \ref{zoom} shows the same data as Figure \ref{error}, but for low intensities on a linear scale.

Without additional noise ($F_{noise}=0$, left image of Figure \ref{zoom}) the difference in $E$ between the systems is small, especially compared to the situation including additional noise (right image of Figure \ref{zoom}). 


It is observed that for $\langle  I\rangle   \lesssim 0.05$ the relative error $E$ of AGIPD is lower than the relative error determined for RAMSES and large regions of interest independently of the noise. While the significance of this result is debatable for $F_{noise}=0$, the result is statistically significant when additional noise is present. It should be noted that the higher signal to noise ratio of AGIPD does not translate directly to a lower relative error.

\section{Summary and discussion}

A simulation tool using the detector simulation program HORUS was set up in order to simulate XPCS experiments.
The numeric simulations took the whole chain into account: modeling a simple real space system, generating diffraction images, finding the detector response and evaluating the data. It was shown that the simulated diffraction patterns show the counting statistics of fully coherent speckle patterns and have a well defined speckle size.

The well known intensity autocorrelation function was used as a method for data evaluation and the most relevant parameters were extracted by a simple exponential fit to the $g_2$ function. The simulations were performed for a set of different detector systems and a set of different noise sources.


A decrease of optical contrast was observed for the non ideal systems and attributed to charge sharing effects. 

The simulated intensity fluctuations (see section \ref{image}) had a very small effect on the data evaluation, which proved the robustness of the intensity autocorrelation technique and its applicability for experiments at free electron laser sources. 

Contrary to common expectation (see Equation \ref{snr_eq}) the signal to noise ratio was found to saturate above intensities larger than approximately 1 photon per (100~$\upmu$m)$^2$. Deviations from the analytical expression were already observed more than one order of magnitude below this intensity.

The signal to noise ratio observed for the AGIPD showed an unexpected transition behavior. At low intensities ($\langle  I\rangle   \lesssim 0.1$) it is similar the signal to noise ratio of RAMSES and small regions of interest (limited by pixel density). At large intensities ($\langle  I\rangle   \gtrsim 1.0$) it is similar to the signal to noise ratio of RAMSES and large regions of interest (limited by total sensitive area). 

The error in the relevant determined physical parameters was shown to deviate in behavior from the signal to noise ratio. The relative error saturated as intensity increased, but at intensities above approximately 0.1 photon per (100~$\upmu$m)$^2$. This is about one order of magnitude lower than the saturation of the signal to noise ratio.

\begin{table}
	\centering
		\begin{tabular}{>{\centering\arraybackslash}m{0.12\textwidth}|>{\centering\arraybackslash}m{0.26\textwidth}|>{\centering\arraybackslash}m{0.26\textwidth}|>{\centering\arraybackslash}m{0.26\textwidth}}
														& \textbf{MAAT} & \textbf{RAMSES lROI} & \textbf{RAMSES sROI} \\
			\noalign{\smallskip}\hline\noalign{\smallskip}
			\textbf{key features}					& effective (100~$\upmu$m)$^2$ pixel size due to aperturing
														& (100~$\upmu$m)$^2$ system with the same number of pixels
														& (100~$\upmu$m)$^2$ system with the same area \\ 
			\noalign{\smallskip}\hline\noalign{\smallskip}
			\textbf{contrast}							& larger, reproducing ideal (100~$\upmu$m)$^2$ system
														& larger, but lower than ideal (100~$\upmu$m)$^2$ system for $\langle  I\rangle  \gtrsim 0.1$ due to charge sharing
														& like RAMSES lROI \\ 
			\noalign{\smallskip}\hline\noalign{\smallskip}											
			\textbf{signal to noise ratio}	& lower for $\langle  I\rangle   \lesssim 1.0$, slightly larger otherwise 
														& like MAAT
														& lower for $\langle  I\rangle   \lesssim 0.05$, larger otherwise \\
			\noalign{\smallskip}\hline\noalign{\smallskip}											
			\textbf{relative error $E$}		& larger for $\langle  I\rangle   \lesssim 0.1$, about the same otherwise 
														& larger for $\langle  I\rangle   \lesssim 0.05$, although hardly significant without additional noise, lower at larger $\langle  I\rangle  $
														& lower except for the two lowest intensities, although hardly significant there
		\end{tabular}
		\caption{Performance results presented in this study in comparison to the AGIPD system. Intensities are given as photons per (100~$\upmu$m)$^2$ area.\label{summary}}
\end{table}

A comparison of the performance of the detector systems is summarized in Table \ref{summary}. The results for the signal to noise ratio of MAAT were partially anticipated in \cite{mid}. Although the systems are quite different in design they vary in the signal to noise ratio only by a factor of 2-3, and even less in the relative error. However the dependence on intensity shows distinctively different features for the different systems. 






The amount of simulated data (5 pulse trains) limited the precision of the results at the lowest intensities. Additional constraints were imposed by the limited number of acquired frames per pulse train. The systematic influences of this proved to be a restricting factor. The finding that the number of frames influences the precision is in good agreement with theoretical predictions (see. e.g. Equation (18) and (19) in \cite{lumma}) and is based on the fact that the $g_2$ function measures pairs of intensities (i.e. correlations). Obviously, the more number of pairs (i.e. frames) are available in an experiment the better the statistics of the measurement.

The simulation was based on the assumption that all detector systems provide the same number of stored frames in a pulse train. In reality a detector system with smaller pixels may lead to a reduced number of stored images. This is affecting both the time scale that can be covered in an XPCS experiment and the SNR in the limit of low
intensities.

In the simulations the sample dynamics were completely independent of the illumination process. In experiments some of the incoming photons will always be absorbed by the sample and its suspending medium, giving rise to a temperature increase and possible changes in the sample. These effects were completely neglected in the simulations presented here, but might ultimately be the dominating effect for experiments at the European XFEL. 

Furthermore, in experimental situations many more complications may occur, e.g. partial coherence and anisotropic speckle sizes, which have been neglected here. The impact of those effects is beyond the scope of this study, but might influence the results for experiments at the European XFEL. 

XPCS experiments require the positional jitter of the illuminating beam to be small compared to the beam size. It has been recently demonstrated at the Linear Coherent Light Source (LCLS) at Stanford, USA, that a jitter between 4\% and 20\% of the beam size can be achieved \cite{LCLS}, thus giving confidence that postional jitter will not seriously influence XPCS experiments at the European XFEL.

\section{Conclusions}

The intuitive conclusion that aperturing is not beneficial as data is 'thrown away' was proven to be correct for low intensities (see Table \ref{summary}). For intensities larger than approximately 1 photon per (100~$\upmu$m)$^2$ aperturing was found to be beneficial, as charge sharing effects were excluded by it.

It was shown that for the investigated case (100~$\upmu$m)$^2$ pixels produced significantly lower relative errors on the extracted correlation times than (200~$\upmu$m)$^2$ pixels when the average intensity exceeded approximately 0.05 photons per (100~$\upmu$m)$^2$.

The signal to noise ratio of the investigated (100~$\upmu$m)$^2$ pixel systems exceeded the signal to noise ratio of the AGIPD when a) the average intensity exceeded approximately 0.05 photons per (100~$\upmu$m)$^2$ and the same area is covered by both systems or b) average intensity exceeded approximately 1 photons per (100~$\upmu$m)$^2$ and both pixels have the same number of pixels.

It was found that under certain conditions the maximum number of frames limited the precision. Thus for XPCS experiments at the European XFEL maximizing the number of stored frames per pulse train might be as important as minimizing the pixel size, which are conflicting requirements. The use of alternate (e.g. logarithmic) sampling schemes to compensate for this limitation will be investigated in a future study.



\section{Acknowledgments}

The authors would like to thank the European XFEL for funding of the development of AGIPD, as well as Anders Madsen from the European XFEL and Fabian Westermeier from the XPCS group at DESY for helping the corresponding author understand the XPCS technique. The simulations presented in this study use programming code that is based in parts on the HORUS program developed by Guillaume Potdevin \cite{AGIPD3, horus1}, now at TU Munich.


\begin{thebibliography}{[1]}
\bibitem{XFEL}{M. Altarelli et al., European X-ray Free Electron Laser. Technical Design Report, ISBN 978-3-935702-17-1 (2006).}
\bibitem{AGIPD1}{B. Henrich et. al., The adaptive gain integrating pixel detector AGIPD a detector for the European XFEL, Nucl. Instr. and Meth. A, DOI: 10.1016/j.nima.2010.06.107.}
\bibitem{AGIPD2}{X. Shi et. al., Challenges in chip design for the AGIPD detector, Nucl. Instr. and Meth. A 624(2) 2010 387-391, DOI: 10.1016/j.nima.2010.05.038.}
\bibitem{AGIPD3}{G. Potdevin et. al., Performance simulation of a detector for 4th generation photon sources: The AGIPD, Nucl. Instr. and Meth. A 607(1) 2009 51-54, DOI: 10.1016/j.nima.2009.03.121.}
\bibitem{grubel}{G. Grubel, F. Zontone, Correlation spectroscopy with coherent X-rays, J. Alloys and Compounds, DOI: 10.1016/S0925-8388(03)00555-3.}
\bibitem{split_delay1}{G. Grübel, G. B. Stephenson, C. Gutt, H. Sinn, Th. Tschentscher, , XPCS at the European X-ray free electron laser facility
Nucl. Instr. and Meth. B 262(2) 2007 357-367, DOI: 10.1016/j.nimb.2007.05.015.}
\bibitem{split_delay2}{W. Roseker et. al., Performance of a picosecond x-ray delay line unit at 8.39 keV, Optics Letters 34 2009 1768-1770.}
\bibitem{gutt}{C. Gutt et. al., Measuring temporal speckle correlations at ultrafast x-ray sources, Optics Express 17(1) 2009 55-61.}
\bibitem{jakeman}{E. Jakemen, in {\it Photon Correlation and Light Beating Spectroscopy}, Ch. 4, p. 75-150, H. Z. Cummins \& E.R. Pike (Eds.), (Plenum, 1973).}
\bibitem{falus}{P. Falus, L. B. Lurio, S. G. J. Mochrie, Optimizing the signal-to-noise ratio for X-ray photon correlation spectroscopy, J. Syn. Rad. 13(3) 2006 253-259, DOI:10.1107/S0909049506006789.}
\bibitem{hansen}{K. Hansen, M. Randall, S. Schleitzer, C. Gutt, System-level simulation of a X-ray imager with nonlinear gain and per-pixel digitizer: XPCS case study, Nucl. Instr. and Meth. A 613(2) 2010 323-333, DOI: 10.1016/j.nima.2009.11.056.}
\bibitem{xpcs_scm_book}{G. Gr\"ubel, A. Madsen, and A. Robert in {\it Soft-Matter Characterization}, Ch. 18, p. 953-995, R. Borsali \& R. Pecora (Eds.), (Springer, 2008).}
\bibitem{fel_fluct}{E. L. Saldin, E. A. Schneidmiller, M. V. Yurkov, Statistical properties of radiation from SASE FEL driven by short electron bunches, Nucl. Instr. and Meth. A 507(1-2) 2003 101-105, DOI: 10.1016/S0168-9002(03)00847-7.}
\bibitem{fowler}{R. F. Fowler, J. V. Ashby, C. Greenough, Computational modelling of semiconducting X-ray detectors, Nucl. Instr. and Meth. A 477(1-3) 2002 226-231, DOI: 10.1016/S0168-9002(01)01878-2.}
\bibitem{droplet}{Y. Chushkin, C. Caronna, A. Madsen, X-ray photon correlation spectroscopy of molecular dynamics using an event correlation scheme. Unpublished (2011).}
\bibitem{sse}{J. Becker, E. Fretwurst, R. Klanner, Measurements of charge carrier mobilities and drift velocity saturation in bulk silicon of $<$111$>$ and $<$100$>$ crystal orientation at high electric fields, Solid-State Electronics, 56(1) 2011 104-110, DOI: 10.1016/j.sse.2010.10.009.}
\bibitem{horus1}{G. Potdevin, U. Trunk, H. Graafsma, HORUS, an HPAD X-ray detector simulation program, J. Inst. 4 2009 P09010, DOI: 10.1088/1748-0221/4/09/P09010.}
\bibitem{david1}{D. Pennicard, et. al., Simulations of charge summing and threshold dispersion effects in Medipix3Nucl. Instr. and Meth. A 636(1) 74-81, DOI: 10.1016/j.nima.2011.01.124.}
\bibitem{david2}{D. Pennicard and H. Graafsma, Simulated performance of high-Z detectors with Medipix3 readout, J. Iinst. 6 2011 P06007, DOI: 10.1088/1748-0221/6/06/P06007.}
\bibitem{horus3}{G. Potdevin, H. Graafsma, Analysis of the expected AGIPD detector performance parameters for the European X-ray free electron laser, Nucl. Instr. and Meth. A, DOI: 10.1016/j.nima.2011.09.012.}
\bibitem{plasma}{J. Becker, D. Eckstein, R. Klanner, G. Steinbruck, Impact of plasma effects on the performance of silicon sensors at an X-ray FEL, Nucl. Instr. and Meth. A 615(2) 2010 230-236, DOI: 10.1016/j.nima.2010.01.082.}
\bibitem{thesis}{J. Becker, Signal development in silicon sensors used for radiation detection, PhD thesis, Hamburg University, July 2010, DESY-THESIS-2010-033.}
\bibitem{plasma2}{J Becker, K Gärtner, R. Klanner, R. Richter, Simulation and experimental study of plasma effects in planar silicon sensors, Nucl. Instr. and Meth. A 624(3) 2010 716-727, DOI: 10.1016/j.nima.2010.10.010.}
\bibitem{schaetzel}{K. Schatzel, Noise on Photon correlation data: I. Autocorrelation functions, Quantum Opt. 2 (1990) 287-305.}
\bibitem{jumpdiff1}{L. N. Gergidis, D. N. Theodorou and H. Jobic, Dynamics of n-Butane-Methane Mixtures in Silicalite, Using Quasielastic Neutron Scattering and Molecular Dynamics Simulations, J. Phys. Chem. B 104(23) 2000, 5541-5552, DOI: 10.1021/jp0000073.}
\bibitem{jumpdiff2}{C. T. Chudley and R. J. Elliott, Neutron Scattering from a Liquid on a Jump Diffusion Model, Proc. Phys. Soc. 77(2) 1961 353-361, DOI: 10.1088/0370-1328/77/2/319.}
\bibitem{mid}{G. Grubel et. al., Report of Working Group II on X-ray Photon Correlation Spectroscopy, 2010, available at \url{http://www.xfel.eu/research/instruments/mid}}
\bibitem{lumma}{D. Lumma, L.B. Lurio, S.G.J. Morchie, M. Sutton, Area detector based photon correlation in the regime of short data batches: Data reduction for dynamic x-ray scattering, Rev. Sci. Instr. 71, 3274 (2000), DOI: 10.1063/1.1287637}
\bibitem{LCLS}{J. L. Turner et. al., FEL beam stability in the LCLS, Proceedings of 2011 Particle Accelerator Conference THP168, \url{http://www.c-ad.bnl.gov/pac2011/proceedings/papers/thp168.pdf}.}

\end{thebibliography}
\end{document}